\documentclass[prl,twocolumn,preprintnumbers,amsmath,amssymb]{revtex4}

\usepackage{graphicx}% Include figure files
\usepackage{dcolumn}% Align table columns on decimal point
\usepackage{bm}% bold math
\usepackage{color}
\usepackage{epstopdf}

\usepackage{ulem}
\def\eps{\varepsilon}

\newcommand{\br}{ {\bm r}}

\def\EE{{\mathbb{E}}}

\def\eps{\varepsilon}

\def\bx{{\bm x}}

\def\br{{\bm r}}

\newtheorem{proposition}{Proposition}[section]

\begin{document}

%\widetext

%\preprint{APS/123-QED}

%\title{Thermalization of random waves in the presence of strong disorder:\\
%Wave turbulence theory, simulations, and experiments in multimode optical fibers(hopefully...)}% Force line breaks with \\
%\title{Nonequilibrium condensation against strong disorder}% Force line breaks with \\
\title{Interplay of thermalization and strong disorder:
Wave turbulence theory, numerical simulations, and experiments in multimode optical fibers}
%\title{Thermalization against equilibration mediated by strong disorder:
%Wave turbulence theory, simulations, and experiments in multimode optical fibers}% Force 
\author{Nicolas Berti$^{1}$, Kilian Baudin$^{1}$, Adrien Fusaro$^{2}$, Guy Millot$^{1,3}$, Antonio Picozzi$^{1}$, Josselin Garnier$^{4}$}
%\author{Josselin Garnier$^{1}$, Kilian Baudin$^{2}$, Adrien Fusaro$^{2}$, Antonio Picozzi$^{2}$}
%\affiliation{$^{1}$ labs}
\affiliation{$^{1}$ Laboratoire Interdisciplinaire Carnot de Bourgogne, CNRS, Universit\'e Bourgogne Franche-Comt\'e, Dijon, France}
\affiliation{$^{2}$ CEA, DAM, DIF, F-91297 Arpajon Cedex, France} 
\affiliation{$^{3}$ Institut Universitaire de France (IUF), 1 rue Descartes, 75005 Paris, France}
\affiliation{$^{4}$ CMAP, CNRS, Ecole Polytechnique, Institut Polytechnique de Paris, 91128 Palaiseau Cedex, France}
%\affiliation{$^{3}$ 
%Institut f\"ur Astrophysik, Georg-August Universit\"at, Friedrich-Hund-Platz 1, D-37077 G\"ottingen, Germany}


\begin{abstract}
We address the problem of thermalization in the presence of a time-dependent disorder 
in the framework of the nonlinear Schr\"odinger (or Gross-Pitaevskii) equation with a random potential.
The thermalization to the Rayleigh-Jeans distribution is driven by the nonlinearity.
On the other hand, the structural disorder is responsible for a relaxation toward the homogeneous equilibrium distribution (particle equipartition), which thus inhibits thermalization (energy equipartition).
On the basis of the wave turbulence theory, we derive a kinetic equation that accounts for the presence of strong disorder.
The theory unveils the interplay of disorder and nonlinearity.
It unexpectedly reveals that a non-equilibrium process of condensation and thermalization can take place in the regime where disorder effects dominate over nonlinear effects.
We validate the theory by numerical simulations of the nonlinear Schr\"odinger equation and the derived kinetic equation, which are found in quantitative agreement without using adjustable parameters.   
Experiments realized in multimode optical fibers with an applied external stress evidence the process of thermalization in the presence of strong disorder.
\end{abstract}

%\pacs{42.65.Sf, 05.45.a}

\maketitle

{\it Introduction.-}
A non-integrable Hamiltonian system of random waves is expected to exhibit a process of thermalization, which is characterized by an irreversible evolution toward the thermodynamic equilibrium state of maximum entropy.
% that realizes the maximum of entropy.
%as a result of an irreversible process of diffusion in phase
In the weakly nonlinear regime, this process is described in detail by the well-developed wave turbulence theory \cite{zakharov92,Newell01,nazarenko11,Newell_Rumpf,shrira_nazarenko13,
laurie12,Lvov10,onorato15,PR14}.
In spite of the formal reversibility of the Hamiltonian system, the wave turbulence kinetic equation describes the actual irreversible evolution to the Rayleigh-Jeans (RJ) equilibrium distribution.
% as expressed by a $H-$theorem of entropy growth.
%which is expressed by a $H-$theorem of entropy growth, in analogy with the celebrated Boltzmann's $H-$theorem in kinetic gas theory.
%A prominent manifestation of RJ thermalization is the possibility to achieve wave condensation, which is featured by the macroscopic population of the fundamental mode of the system \cite{Newell01,nazarenko11,Newell_Rumpf,
%PRL05,berloff07,PD09,Fleischer,suret,magnons15,nazarenko14,PRL18}.
RJ thermalization can be characterized by a process of wave condensation that is featured by the macroscopic population of the fundamental mode of the system \cite{Newell01,nazarenko11,Newell_Rumpf,PRL05,berloff07,PD09,brachet11,nazarenko14,magnons15,
chiocchetta16,PRL18,bloch21}.
This phenomenon received a recent renewed interest with the discovery of spatial beam cleaning in multimode optical fibers (MMFs) \cite{krupa16,wright16,krupa17}.
Along this line, RJ thermalization and light condensation in MMFs have been discussed \cite{PRA11,PRL19,pod19,PRA19,christodoulides19,kottos20,fan22} and recently observed experimentally \cite{PRL20,EPL21,wise_arxiv,mangini22,pod21}.

%, despite the classical nature of the system. 
%Classical wave condensation finds its origin in the natural thermalization of the wave system toward the %thermodynamic 
%Rayleigh-Jeans equilibrium distribution, whose divergence is responsible for the macroscopic occupation of the fundamental mode of the system \cite{Newell01,nazarenko11,Newell_Rumpf,
%PRL05,zakharov92,shrira_nazarenko13,chiocchetta16,PR14,laurie12}.

On the other hand, a structural disorder of the nonlinear medium is known to deeply affect the coherence properties of the waves.
Understanding the interplay of nonlinearity and disorder is a fundamental problem, in relation with the paradigm of statistical light-mode dynamics, glassy behaviors and complexity science \cite{conti22,delre17,churkin15,segev,cherroret15,nazarenko19,wang20,
cherroret21,cherroret20b}.
Disorder is also known to  impact light propagation in MMFs, a feature relevant to endoscopic imaging \cite{psaltis16,faccio19}, or to study completely integrable Manakov systems \cite{mecozzi12a,mecozzi12b,mumtaz13,xiao14}.
Due to refractive index fluctuations introduced by inherent imperfections and environmental perturbations, 
%such as eccentricity, bending, and twisting, 
a MMF leads to both polarization mixing and random mode coupling \cite{mecozzi12a,mecozzi12b,mumtaz13,xiao14,cao16}.
While polarization random fluctuations, i.e., {\it weak disorder}, have been shown to accelerate the process of beam-cleaning condensation in MMFs \cite{PRL19,PRA19}, so far,  the interplay of {\it strong disorder} (i.e., random coupling among non-degenerate modes) and  thermalization has not yet been considered.
%, neither theoretically, nor numerically.
%In all previous works, a genuine random mode coupling (i.e. among non-degenerate modes) was not considered.

In this Letter we address the problem of thermalization of random waves that propagate in a disordered system by considering the representative example of the Nonlinear Schr\"odinger (NLS), or Gross-Pitaevskii, equation with a time-dependent random potential.
%Random coupling among the modes is expected to prevent condensation, simply because the emergence of the fundamental mode (condensate) should be inhibited by its random coupling with the neighbouring modes.
On the basis of the wave turbulence theory \cite{zakharov92,Newell01,nazarenko11,Newell_Rumpf,shrira_nazarenko13,
laurie12,Lvov10,onorato15,PR14},  we derive 
%for the first time 
a kinetic equation (KE) that accounts for the presence of a time-dependent strong  disorder.
Our theory describes in detail the {\it antagonist impacts of nonlinearity and disorder:}
While strong disorder enforces a relaxation to the homogeneous equilibrium distribution of the modal components (`particle' equipartition, $w_j^{\rm eq} =$const, $w_j$ being the occupation of the $j-$th mode), the nonlinear process of thermalization favours 
%the emergence of a condensed state featured by a
the macroscopic population of the condensed fundamental mode ($w_0 \gg w_j$ for $j \neq 0$).
%, and en energy equipartition among the other modes.
The remarkable result of our work is to show that, despite the dominant strength of disorder, the system can exhibit an unexpected process of {\it non-equilibrium condensation} in the initial evolution stage, while the system eventually relaxes to the homogeneous equilibrium distribution dictated by strong disorder.
%The theory reveals that, aside from the expected inter-modal random coupling that inhibits condensation, there exists an intra-modal random coupling that counter-intuitively favours condensation and RJ thermalization. 
The theory is confirmed by intensive numerical simulations of the NLS equation, which are found in quantitative agreement with the simulations of the derived KE, {\it without using any adjustable parameter}.
We report experiments in MMFs with an applied external stress to control the strength of disorder, which evidences the process of RJ thermalization and condensation in the presence of strong disorder.

%Aside from a recent study that considered a z-independent (i.e. stationary) disordered potential [28], fluctuations of the medium are usually introduced phenomenologically in the final kinetic equation [35]. We propose to develop a systematic method to tackle the impact of a coloured spatio-temporal disorder on the fourth-order moment equations, as well as its role on the regularization of wave resonances and the problem of achieving a closure of the hierarchy of moment equations in wave turbulence. V. M. Malkin and N. J. Fisch, Transition between inverse and direct energy cascades in multiscale optical turbulence, Phys. Rev. E 97 (2018), 032202. 14

\newpage

Our work paves the way for the development of a systematic method to tackle the impact of a {\it time-dependent disorder in wave turbulence} -- our methodology substantially differs  from that developed for a time-independent disorder \cite{cherroret15,nazarenko19,wang20,cherroret21,cherroret20b}.
%Our work paves the way for the development of a systematic multiscale analysis method to tackle the impact of a {\it time-dependent disorder in wave turbulence} -- our methodology substantially differs from that developed for a time-independent disorder [] that is based on diagrammatic expansions.
More generally this work contributes to the understanding of spontaneous organization of coherent states in nonlinear disordered systems \cite{conti22,delre17,churkin15,segev,cherroret15}.
%, in relation with the paradigm of statistical light-mode dynamics 
%glassy behaviors and complexity 
%in random lasers 
%Furthermore, MMFs are important 
%attracting a renewed interest 
%for telecommunication applications 
%in spatial division multiplexing 
%\cite{richardson}.
% and 
%as a platform 
%for novel fiber laser sources \cite{wright17}. 
%Presence of random potential makes our problem essentially different from a standard wave turbulence system in which the only source of randomness is in the initial data. Thus, in standard wave turbulence, the initial randomness is assumed to be propagating over the nonlinear time, whereas in presence of random potentials, it is naturally present at all times. We, therefore, believe our work lays the basis to a new way of studying the above problem, with specific road map assumptions of general interest.

{\it NLS equation with random potential.-}
We consider the general form of the stochastic NLS equation 
%governing the transverse spatial evolution of an optical beam propagating along the $z-$axis:
% of a waveguide modeled by
%a confining potential $V_0(\br)$ [with $\br = (x, y)$].
%light propagation in a nonlinear waveguide (trapping) potential
\begin{equation}
i \partial_z \psi= - \alpha \nabla^2 \psi +V(\br) \psi -\gamma |\psi|^2 \psi + \delta V(\br,z) \psi  .
\label{eq:psi}
\end{equation}
It governs
%, in particular, 
the transverse spatial evolution of an optical beam propagating along the $z-$axis of a waveguide, whose ideal transverse index profile is $V(\br)$ [$\br = (x, y)$], while $\delta V(z,\br)$ is the `time'-dependent random perturbation of the potential 
%(refractive index fluctuations), assumed to be real-valued with 
($\left< \delta V\right>=0$).
The parameters $\alpha$ and $\gamma$ denote the linear and nonlinear coefficients.
The disorder being (`time') $z-$dependent, our system is of different nature than those studying the interplay of thermalization and Anderson localization \cite{cherroret15,nazarenko19,wang20,cherroret21,cherroret20b}.
%We consider the eigenvalues/eigenmodes $(\beta_j,u_j)$ solution of $\beta_j u_j = - \alpha \nabla^2 u_j +V_0(\bx) u_j$.
%We can take the eigenmodes to be real-valued.
%We assume that the $N$ modes are non degenerate, i.e. all $\beta_j$ are distinct.

We expand the field $\psi(z,\br) = \sum_{j} a_j(z) u_j(\br)$ on the basis of the $M$ real-valued eigenmodes $u_j(\br)$ (solution of $\beta_j u_j = - \alpha \nabla^2 u_j +V(\br) u_j$) of the unperturbed waveguide. 
The mode amplitudes $a_j(z)$ satisfy 
\begin{equation}
i\partial_z a_j = \beta_j a_j -\gamma \sum_{l,m,n} Q_{jlmn} a_l a_m a_n^* +\sum_{l}C_{jl}(z) a_l ,
\label{eq:a_j}
\end{equation}
where $Q_{jlmn}=\int u_j(\br) u_l(\br) u_m(\br)  u_n(\br) d\br$ denotes the mode overlap, and the random mode coupling matrix reads
\begin{equation}
C_{jl}(z) = \int u_j(\br) \delta V(z,\br)  u_l(\br) d\br .
\end{equation}
The stochastic NLS Eq.(\ref{eq:psi}) and the modal NLS Eq.(\ref{eq:a_j}) are equivalent.
They conserve the total power (particle number) $N=\int |\psi|^2 d\br = \sum_j |a_j|^2$, while the random potential $\delta V(\br,z)$ in Eq.(\ref{eq:psi}) (or 
%the random matrix 
${\bf C}(z)$ in Eq.(\ref{eq:a_j})), prevents the conservation of the  energy (Hamiltonian).

\begin{center}
\begin{figure}
\includegraphics[width=1\columnwidth]{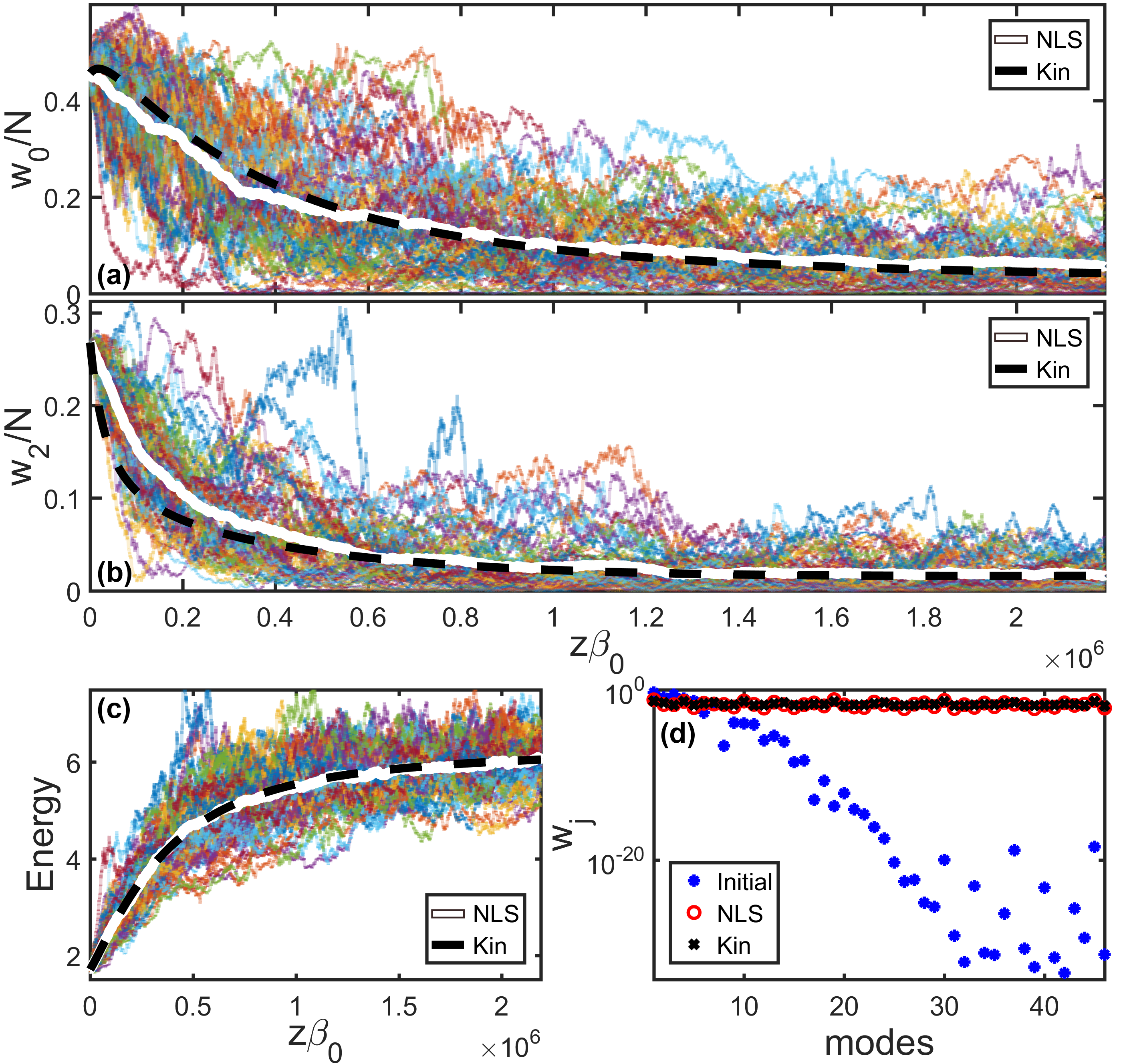}
\caption{
\baselineskip 10pt
{\bf Dynamics dominated by strong disorder ${\cal L}^{\rm RJ}_{\rm kin} \gg {\cal L}^{\rm eq}_{\rm kin}$:} 
The system irreversibly relaxes toward the equilibrium $w_j^{\rm eq}$.
Evolutions of the fundamental mode $w_0(z)$ (a), and $w_2(z)$ (b), the energy $E(z)/(N \beta_0)$ (c), obtained from the numerical simulation of the NLS Eq.(\ref{eq:a_j}): 64 realizations are reported with colored lines; 
the bold white line is the corresponding empirical average;
the dashed black line is the prediction of the KE (\ref{eq:kin}).
(d) Modal distribution $w_j$ in the initial condition ($z=0$, blue) and at $z \beta_0=2 \times 10^6$ for the NLS simulation (red), and the KE (black).
Parameters: $L_{dis}/L_{lin}=7$, $L_{dis}/L_{nl}= 4.1 \times 10^{-4}$, $\ell_c \beta_0=42$.
% ${\cal L}^{\rm RJ}_{\rm kin} / {\cal L}^{\rm eq}_{\rm kin} \simeq 400$. 
%Parameters: 46 modes, $\beta_0 = 8371$m, $\sigma =1200$m$^{-1}$, $L_{NL0} = 2$m, $\ell_c = 5$mm, $b_x = 0.4$, $b_y = 0.5$, 64 realizations, $\beta_{0x} = \sqrt{2} \beta_{0y}$.
%[The parameters in the simulations should be given in term of $\beta_{p_x=0,p_y=0}=\beta_0=(\beta_{0,x}+\beta_{0,y})/2=\sqrt{\alpha}(\sqrt{q_x}+\sqrt{q_y})$: $\sigma/\beta_0=?$, $\beta_0 \ell_c=?$, $\beta_0 L_{nl} = ?$.]
}
\end{figure}
\end{center}

\begin{center}
\begin{figure}
\includegraphics[width=1\columnwidth]{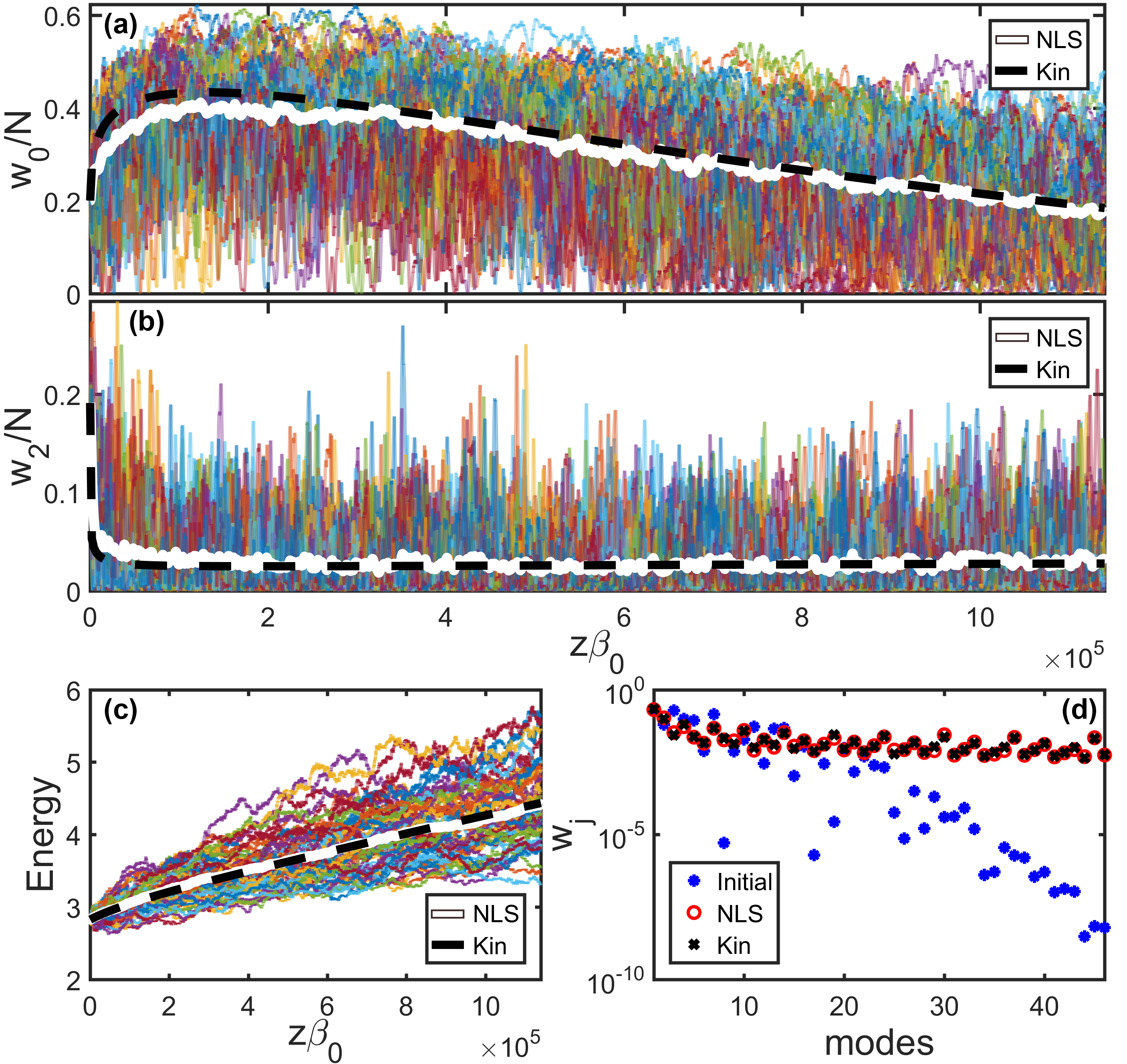}
\caption{
\baselineskip 10pt
{\bf Thermalization precedes equilibrium relaxation:}
Same panels as in Fig.~1, but in the regime ${\cal L}^{\rm RJ}_{\rm kin} \lesssim {\cal L}^{\rm eq}_{\rm kin}$.
The system exhibits an incipient process of RJ thermalization and nonequilibrium condensation characterized by a growth of the condensate amplitude $w_0(z)$ for $z \beta_0 \lesssim 2\times 10^5$.
Disorder subsequently prevails, which induces a decay of $w_0(z)$ (and eventually brings the system to equilibrium $w_j^{\rm eq}$). 
%Evolutions of the fundamental mode $w_0(z)$ (a), and $w_2(z)$ (b), the energy $E(z)/(N \beta_0)$ (c), obtained from the numerical simulation of the NLS Eq.(\ref{eq:a_j}): 64 realizations are reported with colors and the corresponding average with the white line. Simulation of the kinetic Eq.(\ref{eq:kin}) (dashed lack line). (d) Modal distribution $w_j$ in the initial condition ($z=0$, blue) and at $z \beta_0=8 \times 10^5$ for the NLS simulation (red, averaged over the 64 realizations), and the kinetic equation (black).
Parameters: $L_{dis}/L_{lin}=7$, $L_{dis}/L_{nl}= 0.033$, $\ell_c \beta_0=167$.
% ${\cal L}^{\rm RJ}_{\rm kin} / {\cal L}^{\rm eq}_{\rm kin} \simeq 0.07$. 
%Parameters: 46 modes, $\beta_0 = 8371$m, $\sigma =1200$m$^{-1}$, $L_{NL0} = 25$mm, $\ell_c = 20$mm, $b_x = 0.4$, $b_y = 0.5$, 64 realizations, $\beta_{0x} = \sqrt{2} \beta_{0y}$.
}
\end{figure}
\end{center}

\begin{center}
\begin{figure}
\includegraphics[width=1\columnwidth]{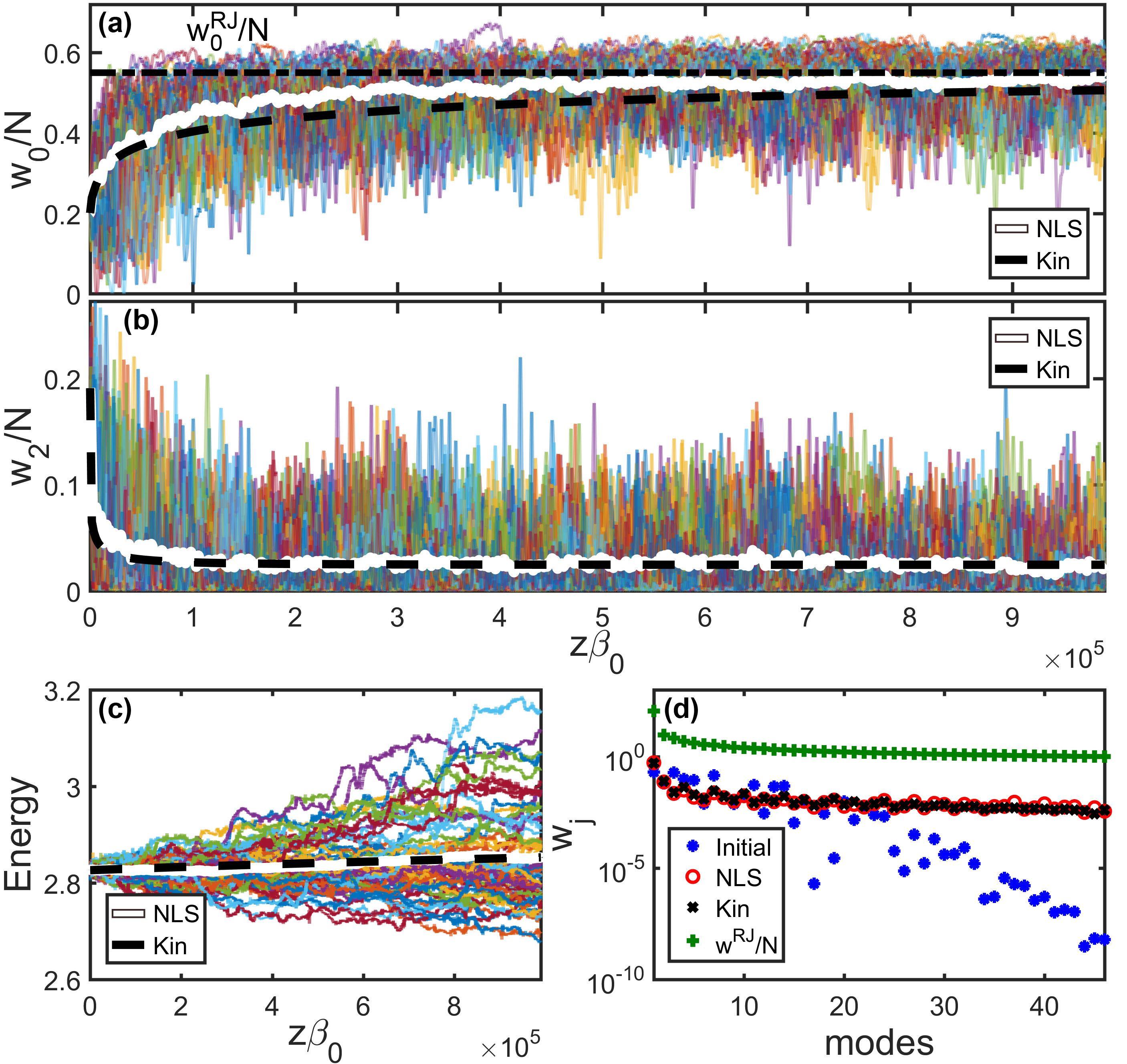}
\caption{
\baselineskip 10pt
{\bf Thermalization prevails over equilibrium relaxation:}
Same panels as in Fig.~1, but in the regime ${\cal L}^{\rm RJ}_{\rm kin} \ll {\cal L}^{\rm eq}_{\rm kin}$.
The system exhibits a process of RJ thermalization and condensation characterized by a significant growth of the condensate amplitude $w_0(z)$ to the value predicted by the  RJ distribution, $w_0^{\rm RJ}/N \simeq 0.55$ (horizontal dashed-dotted black line) (a).
At variance with Figs.~1-2, the energy $E(z)$ is almost constant (c).
The modes approach the RJ distribution $w_j^{\rm RJ}$ (green) (d).
%Disorder subsequently prevails and eventually brings the system to equilibrium $w_j^{\rm eq}$. 
%Evolutions of the fundamental mode $w_0(z)$ (a), and $w_2(z)$ (b), the energy $E(z)/(N \beta_0)$ (c), obtained from the numerical simulation of the NLS Eq.(\ref{eq:a_j}): 64 realizations are reported with colors and the corresponding average with the white line. Simulation of the kinetic Eq.(\ref{eq:kin}) (dashed lack line). (d) Modal distribution $w_j$ in the initial condition ($z=0$, blue) and at $z \beta_0=8 \times 10^5$ for the NLS simulation (red, averaged over the 64 realizations), and the kinetic equation (black).
Parameters: $L_{dis}/L_{lin}=8.4$, $L_{dis}/L_{nl}= 0.04$, $\ell_c \beta_0=4 \times 10^3$.
% ${\cal L}^{\rm RJ}_{\rm kin} / {\cal L}^{\rm eq}_{\rm kin} \simeq 0.003$. 
%Parameters: 46 modes, $\beta_0 = 8371$m, $\sigma =1000$m$^{-1}$, $L_{NL0} = 25$mm, $\ell_c = 0.5$m, $b_x = 0.3$, $b_y = 0.4$, 64 realizations, $\beta_{0x} = \sqrt{2} \beta_{0y}$.
}
\end{figure}
\end{center}

{\it Kinetic equation.-}
We consider the situation where the random potential is a weak perturbation with respect to linear propagation effects ($\delta V \ll V$), i.e. $L_{lin} = 1/\beta_0 \ll L_{dis} =1/\sigma$ and $L_{lin} \ll \ell_c$, 
%where $\beta_0$ is the fundamental mode eigenvalue and 
where $\sigma^2$ denotes the variance of the fluctuations of the random potential (i.e., `strength' of disorder) and $\ell_c$ the corresponding correlation length.
Note that this is the usual case in an optical waveguide configuration, e.g., in MMFs.
Furthermore, we assume that  disorder dominates over nonlinear effects $L_{dis} \ll L_{nl} \simeq 1/(\gamma \left< |\psi|^2\right>)$.

%Combining the (discrete) wave turbulence theory\cite{nazarenko11,Lvov10,bustamante14} with asymptotic analysis of randomly driven  differential equations (time-dependent disorder) \cite{fouque07}, we derive the KE governing the evolution of the averaged modal components $w_j(z)=\left< |a_j(z)|^2\right>$ \cite{supplement}:
%We extend the wave turbulence theory by using tools developed for the asymptotic analysis of randomly driven linear differential equations.  
We develop a wave turbulence theory 
\cite{zakharov92,Newell01,nazarenko11,Newell_Rumpf,shrira_nazarenko13,
laurie12,Lvov10,onorato15,PR14} accounting for a time-dependent disorder by exploiting tools inherited from the asymptotic analysis of randomly driven ordinary differential equations \cite{fouque07}.  
We derive the KE governing the evolution of the averaged modal components $w_j(z)=\left< |a_j(z)|^2\right>$ \cite{supplement}:
% On the basis of the wave turbulence theory and using tools inhererited from asymptotic analysis of randomly driven  differential equations (time-dependent disorder) \cite{fouque07}
\begin{equation}
\partial_z w_j  = 
\sum_{l\neq j} \Gamma_{jl}^{\rm OD} \big(w_l-w_j \big) + {\cal C}oll[{\bm w}]
\label{eq:kin}
\end{equation}
where the collision term reads 
\begin{eqnarray*}
&& {\cal C}oll[{\bm w}]=8 \gamma^2   \sum_{l,m,n}  \frac{\delta_{jlmn}^K  Q_{jlmn}^2}{G_{jlmn}^{\rm D}} {R}_{jlmn}[{\bm w}], \\
&& {R}_{jlmn}[{\bm w}]=w_l w_m w_j + w_l w_m w_n  -  w_j w_n w_m - w_j w_n w_l,
%\label{eq:kin}
\end{eqnarray*}
%with ${\bm M}_{jlmn}=w_l w_m w_j + w_l w_m w_n  -  w_j w_n w_m - w_j w_n w_l$, 
and the Kronecker symbol denotes a frequency resonance ($\delta_{jlmn}^K=1$ if $\Delta \beta_{jlmn}=\beta_j-\beta_l-\beta_m+\beta_n=0$, and zero otherwise).
For clarity, we assume that the modes are not degenerate -- see \cite{supplement} for the KE accounting for mode degeneracies.
%($\beta_l \neq \beta_j$ for all $j \neq l$), see see \cite{supplement} for the extension to mode degeneracies.

The KE (\ref{eq:kin}) unveils the interplay of nonlinearity and disorder.
It reveals that diagonal and off-diagonal elements of the random matrix ${\bf C}$ play fundamental different roles.
The first term in the KE (\ref{eq:kin}) originates in {\it off-diagonal} elements of $C_{jl}$ ($j \neq l$):
\begin{equation}
\Gamma^{\rm OD}_{jl}= 2 \int_0^\infty\left<C_{jl}(0)C_{jl}(z)\right> \cos \big((\beta_j-\beta_l)z\big) dz.
\end{equation}
It describes an irreversible relaxation toward the homogeneous distribution featured by an equipartition of `particles' among the modes, $w_j^{\rm eq}=N/M=$const.
This process occurs over the typical propagation length \cite{comment}
\begin{equation}
{\cal L}^{\rm eq}_{\rm kin}\simeq  1/\overline{\Gamma_{jl}^{\rm OD}}.
\end{equation}
This is the well-known evolution of a system ruled by random mode coupling.

%The main result of our work is to 
We now show that this relaxation process mediated by strong disorder {\it does not necessarily inhibit the nonlinear processes of thermalization and condensation}.
This becomes apparent through the collision term in the KE (\ref{eq:kin}), which exclusively involves the {\it diagonal} components $C_{jj}$:
\begin{equation}
\Gamma^{\rm D}_{jl}= 
 \int_0^\infty \left<C_{jj}(0)C_{ll}(z)\right> + \left<C_{ll}(0)C_{jj}(z)\right> dz.
\end{equation}
The matrix ${\bf \Gamma}^{\rm D}$ contributes to the tensor involved in the collision term, $G_{jlmn}^{\rm D}=\Gamma^{\rm D}_{ll}+\Gamma^{\rm D}_{mm}+\Gamma^{\rm D}_{nn}+\Gamma^{\rm D}_{jj}+2\Gamma^{\rm D}_{lm}-2\Gamma^{\rm D}_{ln}-2\Gamma^{\rm D}_{lj}-2\Gamma^{\rm D}_{mn}-2\Gamma^{\rm D}_{mj}+2\Gamma^{\rm D}_{nj}$ \cite{supplement}.
To discuss the role of the collision term, we forget for a while the first term in the KE (\ref{eq:kin}).
%\blue{Following the usual proof \cite{zakharov92}, it can be shown that} 
The collision term conserves $N=\sum_j w_j(z)$, $E=\sum_j \beta_j w_j(z)$, and exhibits a $H-$theorem of entropy growth $\partial_z S \ge 0$, with $S(z)=\sum_j \log[w_j(z)]$ \cite{supplement}.
Hence, it describes a process of thermalization to the RJ distribution $w_j^{\rm RJ}=T/(\beta_j-\mu)$, 
%featured by an energy equipartition among the modes.
which occurs over a typical propagation length 
\begin{equation}
{\cal L}^{\rm RJ}_{\rm kin} \simeq L_{nl}^2 \overline{G_{jlmn}^{\rm D}/Q_{jlmn}^2 }.
\end{equation}

For an energy smaller than a critical value $E \le E_{\rm crit} \simeq N \beta_0 \sqrt{M/2}$, the RJ distribution $w_j^{\rm RJ}$ exhibits a phase transition to a condensed state \cite{PRL20}.
The condensate amplitude $w_{0}$ then constitutes the natural parameter that distinguishes the two antagonist regimes:\\
(i) For ${\cal L}^{\rm eq}_{\rm kin} \ll {\cal L}^{\rm RJ}_{\rm kin}$, the disorder dominates and 
$w_0(z) \to w_j^{\rm eq}=N/M=$const for $j=0,1,...,M-1$;\\
(ii) For ${\cal L}^{\rm eq}_{\rm kin} \gg {\cal L}^{\rm RJ}_{\rm kin}$, the dynamics is dominated by RJ thermalization, and  condensation leads to a macroscopic population of the fundamental mode $w_0(z) \to w_0^{\rm RJ} \gg w_{j}^{\rm RJ}$ for $j=1,...,M-1$.

%According to the theory, the two terms in the KE (\ref{eq:kin}) are antagonists and compete.
%If ${\cal L}^{\rm eq}_{\rm kin} \ll {\cal L}^{\rm RJ}_{\rm kin}$, then disorder prevails and the system exhibits the expected relaxation to equilibrium $w_j^{\rm eq}$.
%Unexpectedly, however, a {\it nonequilibrium} process of condensation and thermalization can be observed in the initial stage of propagation when ${\cal L}^{\rm eq}_{\rm kin} \gg {\cal L}^{\rm RJ}_{\rm kin}$, while asymptotically the system inevitably relax to the equilibrium state $w_j^{\rm eq}$.

\newpage

{\it Numerical simulations.-}
We have performed numerical simulations to test the validity of our theory.
We have considered the concrete example of a parabolic trapping potential of the form $V(\br)=q_x x^2 + q_y y^2$, with the fundamental mode eigenvalue $\beta_0=\sqrt{\alpha}(\sqrt{q_x}+\sqrt{q_y})$.
%Aside from Bose-Einstein condensates, the parabolic-shaped trapping models graded-index MMFs where 
%spatial beam cleaning \cite{krupa16,wright16,liu16,krupa17}, as well as 
%light thermalization has been recently observed experimentally \cite{PRL20,EPL21,wise_arxiv,mangini22}.
We consider a general model of disorder with a random potential of the form $\delta V(\br,z)=\mu(z) g(\br)$, where $\mu(z)$ is a real-valued stochastic function with zero mean and $\left< \mu(0) \mu(z)\right>= \sigma^2 \exp(-|z|/\ell_c)$.
In order to remove mode degeneracies, we consider in the simulations an elliptical parabolic potential ($q_x \neq q_y$). 
To compute the matrices ${\bf \Gamma}^{\rm OD}$ and ${\bf \Gamma}^{\rm D}$ in analytical form we consider
%the random mode coupling 
$g(x,y)=\cos(b_x x/x_{0})\cos(b_y y/y_{0})$, where $(x_{0},y_{0})$ denote the radii of the fundamental elliptical mode \cite{supplement}.

According to the theory, the two terms in the KE (\ref{eq:kin}) are antagonists and compete against each other.
If ${\cal L}^{\rm eq}_{\rm kin} \ll {\cal L}^{\rm RJ}_{\rm kin}$, disorder prevails and the system relaxes to the expected equilibrium $w_j^{\rm eq}=$const.
This is illustrated in Fig.~1, which reports the results of the numerical integration of the NLS Eq.(\ref{eq:a_j}) for 64 realizations (${\cal L}^{\rm RJ}_{\rm kin} / {\cal L}^{\rm eq}_{\rm kin} \simeq 400$).
The corresponding average over such realizations (bold white line) is in agreement with the  simulation of the KE (\ref{eq:kin}) (dashed black line) starting from the same initial condition.
Here and thereafter, the quantitative agreement between NLS and KE simulations is obtained without any adjustable parameter.

Unexpectedly, however, a {\it nonequilibrium} process of condensation and thermalization can be observed in the initial stage of propagation when ${\cal L}^{\rm RJ}_{\rm kin} \lesssim {\cal L}^{\rm eq}_{\rm kin}$ (see Fig.~2 for ${\cal L}^{\rm RJ}_{\rm kin} / {\cal L}^{\rm eq}_{\rm kin} \simeq 0.07$), while asymptotically the system still  relaxes to the homogeneous equilibrium state $w_j^{\rm eq}$.
The nonequilibrium property of condensation is reflected by the fact that the energy $E(z)=\sum_j \beta_j w_j(z)$ {\it is not conserved during the evolution}, see Fig.~2(c).
%Note that the parameters in Fig.~2 verify $L_{dis} \sim \sqrt{L_{nl} L_{lin}}$.
%\blue{Note that, the introduction of losses, distributed either homogeneously or non-homogeneously amongst the modes, does not significantly affect the condensate peak in Fig.~2 \cite{supplement}.}

We stress that the condensation processes can occur very efficiently by increasing the correlation length $\ell_c$, in such a way that ${\cal L}^{\rm RJ}_{\rm kin} \ll {\cal L}^{\rm eq}_{\rm kin}$, see Fig.~3 for ${\cal L}^{\rm RJ}_{\rm kin} / {\cal L}^{\rm eq}_{\rm kin} \simeq 0.003$. 
In this regime, the energy is almost conserved $E \simeq$~const and RJ thermalization occurs almost completely, as confirmed by the modal populations that approach the RJ distribution $w_j^{\rm RJ}$ (Fig.~3(d)), and the condensate approaches the RJ prediction $w_0^{\rm RJ}/N \simeq 0.55$, see Fig.~3(a).
% in the absence of random mode coupling (disorder).
Note that, for $z \gg {\cal L}^{\rm eq}_{\rm kin}$, the system would still relax to 
the equilibrium $w_j^{\rm eq}$.
% imposed by strong disorder.

%\blue{However, it is important to remark that, although random mode coupling does not affect the RJ equilibrium, it significantly accelerates the rate of thermalization to this equilibrium.
%Indeed, the ratio of thermalization in the presence and the absence of disorder scales as $\sim {\bar G}^{\rm D}/ \beta_0 \ll 1$.
%This effect of acceleration of thermalization is apparently similar to that pointed out in Ref.\cite{PRL19,PRA19}.
%However, in marked contrast with the previous works \cite{PRL19,PRA19} where only random polarization fluctuations within each single mode were considered (i.e., weak disorder), here we consider random mode coupling (i.e. strong disorder) among all the modes.}

\begin{center}
\begin{figure}
\includegraphics[width=1\columnwidth]{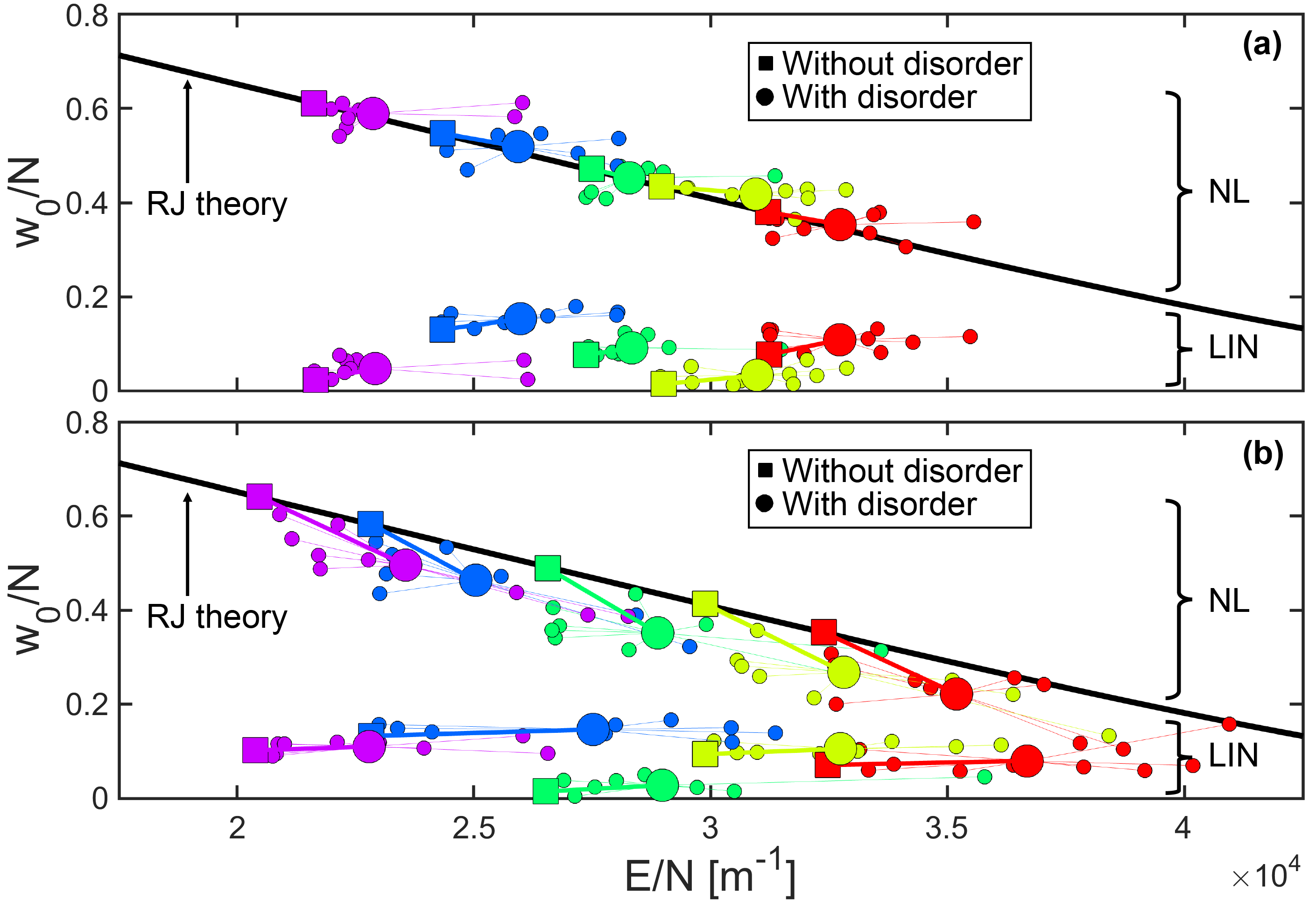}
\caption{
\baselineskip 10pt
{\bf Observation of light condensation with strong disorder:} 
Measurements of the condensate fraction $w_0/N$ vs $E/N$ at small power [linear (LIN) regime] and high power [nonlinear (NL) regime], for a moderate (a), and a large (b), strength of random mode coupling. 
The solid line reports the prediction from the RJ theory, $w_0^{\rm RJ}/N$.
In the absence of strong disorder (squares): $w_0/N$ increases as the power increases, and reaches the RJ prediction (solid line) -- each color refers to a different value of $E/N$.
In the presence of strong disorder (big circles): the energy $E/N$ increases due to disorder (squares are shifted to big circles of the same color), by $\Delta \overline{E/N} \simeq 6$\% (a), and  $\Delta \overline{E/N} \simeq 11$\% (b).
%The small circles report 10 different realizations of disorder for each color, the big circles report the corresponding average over realizations.
The big circles report the average over 10 different realizations of disorder (10 small circles for each color).
RJ thermalization takes place in the presence of strong disorder (a), and it is quenched by further increasing the amount of disorder (b) \cite{supplement}.
}
\end{figure}
\end{center}

{\it Experiments.-}
We performed experiments in a MMF to evidence light condensation in the presence of strong disorder.
The subnanosecond pulses delivered by a Nd:YAG laser ($\lambda=1.06\mu$m) are passed through a diffuser before injection into a 12m long graded-index MMF (i.e., parabolic-shaped potential $V(\br)$) that guides $M \simeq 120$ modes.
We measure the power $N$ and the energy $E$ from the near-field and far-field measurements of the intensity distributions at the fiber output, see Ref.\cite{PRL20} for details.

Here, the originality with respect to {\it all previous experiments} on beam cleaning condensation and thermalization \cite{krupa16,wright16,krupa17,PRL20,EPL21,wise_arxiv,mangini22}, is that we introduce strong disorder in the experiment. 
Strong mode coupling is obtained by applying a stress to the MMF with clamps \cite{cao16}.
By adjusting the applied stress, we can tune the strength of mode coupling (i.e., $\sigma$).
In the absence of an applied stress, polarization coupling 
%within modes, 
and random coupling among degenerate modes take place: In this {\it weak random coupling regime the energy $E=\sum_j \beta_j w_j$ is conserved} during light propagation in the MMF \cite{PRL19,PRA19}, as confirmed by direct experimental measurements \cite{PRL20,wise_arxiv,pod19,EPL21,mangini22}.
Here, we apply stress on the MMF  to induce a random coupling among 
%groups of 
{\it non-degenerate modes} \cite{kahn14,mecozzi12a,mecozzi12b,mumtaz13,xiao14}.
%Here, we use clamps to induce a stress on the MMF, in such a way that {\it groups of non-degenerate modes get randomly coupled} \cite{kahn14,mecozzi12a,mecozzi12b,mumtaz13,xiao14}.
In this regime of {\it strong mode coupling}, the energy $E$ is no longer conserved through propagation in the MMF.
Note that, it would be difficult, or even impossible, to accurately model in the simulations the peculiar form of disorder induced by the applied stress on the fiber.
Furthermore, the simulations reported above do not account for the mode degeneracies of the fiber used in the experiments.
Accordingly, the simulations do not describe quantitatively our experiments.
%Accordingly, the simulations reported above that neglect mode degeneracies do not provide a quantitative description of the experiments.
%Our aim here, is to provide a phenomenological demonstration that condensation does occur in the presence of strong disorder.}

%The energy E provides a measure of the ?coherence? of the beam in the sense that E increases as the beam populates higher-order modes: by increasing the coherence, E decreases and this leads to an increase of the condensate amplitude n0 after nonlinear propagation in the MMF, as described by the condensation curve in Fig. 2(c).
%We recall that the energy reflects the degree of incoherence of the beam, i.e., the larger the number of modes populated at the fiber input, the higher the value of $E$.

We report in Fig.~4(a) the measurements of the condensate fraction $w_0/N$ at small power (linear regime), strong power (nonlinear regime), and in the presence, or absence, of applied stress.
Following Ref.\cite{PRL20}, 
%Following the procedure of Ref.\cite{PRL20}, 
the diffuser allows us to vary the energy density $E/N$ of the injected speckle beam: The squares in Fig.~4(a) report the corresponding condensate fractions $w_0/N$ in the linear and nonlinear regimes in the {\it absence of strong disorder}. %In the absence of any applied stress, $E$ is conserved and the ?? points in Fig.~4 denote , either at small or strong power.
%The MMF is wound on a drum, and 
%For a given speckle beam with energy $E$, we record different realizations of disorder by applying stress with the clamps on different positions of the MMF.
%The detailed procedure is described in \cite{supplement}.
%In substance, 
For each speckle beam with energy $E/N$ (i.e., for each color in Fig.~4(a)), we apply stress on different points of the fiber to get an ensemble of 10 realizations with disorder.
We report in Fig.~4(a) the corresponding values of $w_0/N$ for such 10 realizations (small circles), as well as the corresponding average over realizations (large circles).
%the average over such an ensemble of realizations in Fig.~4 .
% which confirms that strong mode coupling leads to an {\it increase} of the energy.
Because the applied stress on the MMF induces power losses (10\% in Fig.~4(a)) \cite{cao16}, we normalize the energy with respect to the (average) power: $E/N=\sum_j \beta_j w_j/\sum_j w_j = \overline{\beta_j}$.
% which denotes the average of the energy levels $\overline{\beta_j}$ populated by the  distribution $w_j$.
Strong random mode coupling leads 
%to the population of higher-order modes, which leads 
to an {\it increase} of $E/N$, as evidenced in Fig.~4(a) where the squares are shifted to the big circles by an amount of $\Delta \overline{E/N} \simeq 6$\%: The larger the strength of applied stress, the larger the energy shift $\Delta \overline{E/N}$.
%Interestingly, the increase of $E/N$ due to disorder is of the same order in the linear regime (small power), and nonlinear regime (strong power).

Figure~4(a) remarkably reveals that, by increasing the power from the linear regime ($N=0.23$kW) to the nonlinear regime ($N=7$kW), the condensate fraction $w_0/N$ (big circles) increases and approaches the value predicted by the RJ distribution $w_0^{\rm RJ}/N$ (solid line).
Thermalization then takes place: (i) in the presence of strong disorder, i.e., in the presence of an energy shift $\Delta \overline{E/N}$; (ii) over a broad range of $E/N$, i.e., broad range of condensate fractions.
Note that, the presence of losses, distributed either homogeneously or non-homogeneously among the modes, does not significantly affect the condensate fraction \cite{supplement}.

We have repeated the procedure of Fig.~4(a) by increasing the applied stress on the MMF with an energy shift $\Delta \overline{E/N} \simeq 11$\% (20\% of power losses).
As evidenced in Fig.~4(b), $\Delta \overline{E/N}$ is larger than in Fig.~4(a). 
%The results reported in Fig.~4(b) evidence that the increment of $E/N$ due to disorder is larger than in Fig.~4(a).
Consequently, the condensate fractions $w_0/N$ in the nonlinear regime no longer reach the  RJ prediction, i.e., strong disorder prevents a complete process of RJ thermalization and condensation.
By further increasing the applied stress and the corresponding energy shift $\Delta \overline{E/N} \simeq 19$\%, our experimental results show that {\it RJ thermalization is inhibited by strong disorder}, see \cite{supplement}.

%By gradually adjusting the stress, we have realized a transition from weak to strong mode coupling, which corresponds to the transition from single scattering to multiple scattering in mode space.
%We apply stress to the fiber with clamps to enhance the mode coupling. 
%To evaluate the strength of mode coupling in the fiber, the transmission matrix is transformed to the mode basis by decomposing the input and output fields by LP modes

{\it Conclusion.-}
We have developed a wave turbulence theory that accounts for a `time'-dependent disorder by considering the NLS equation with a random potential.
%The theory remarkably reveals that RJ thermalization and condensation can take place even when the dynamics is dominated by strong disorder ($L_{dis}/L_{nl} \ll 1$), provided that $L^{ \rm RJ} \ll L_{eq}$.
%Experiments in MMFs confirm, at a qualitative level, that RJ thermalization and condensation do occur in the presence of strong disorder (strong mode coupling).
%At variance with weak disorder, strong disorder introduces very large fluctuations of the modal components $w_j$ (see NLS simulations in Figs.~1-3), whose average over the realizations is found in quantitative agreement with the simulations of the kinetic Eq.(\ref{eq:kin}), {\it without using any adjustable parameter}.
Simulations of the derived KE (\ref{eq:kin}) are found in quantitative agreement with NLS simulations, {\it without using any adjustable parameter}.
The theory remarkably reveals that RJ thermalization and condensation can take place efficiently in the presence of strong disorder, as confirmed by experiments realized in MMFs.

%Our theoretical approach paves the road to tackle the impact of a `time'-dependent disorder in wave turbulence.
The developed wave turbulence theory can be extended to dissipative systems \cite{fischer10,turitsyn13}, or to different types of disordered nonlinear systems, e.g., Bose-Einstein condensates, hydrodynamics, condensed matter, etc.

%From a broader perspective, the spatiotemporal kinetic formulation we develop here paves the way to the study of novel forms of global incoherent collective behaviors in wave turbulence,

{\it Acknowledgments.-} The authors are grateful to K. Krupa and S. Rica for fruitful discussions. Fundings: Centre national de la recherche scientifique (CNRS), Conseil r\'egional de Bourgogne Franche-Comt\'e, iXCore Research Fondation, Agence Nationale de la Recherche (ANR-19-CE46-0007, ANR-15-IDEX-0003, ANR-21-ESRE-0040). Calculations were performed using HPC resources from DNUM CCUB (Centre de Calcul, Universit\'e de Bourgogne).
%H2020 Marie Sklodowska-Curie Actions (MSCA-COFUND) (MULTIPLY Project No. 713694).

%\begin{widetext}

%%%%%%%%%%%%%%%%%%%%%%%%%%%%%%%%%%%%%%%%%%%%%%%%%
%%%%%%%%%%%%%%%%%%%%%%%%%%%%%%%%%%%%%%%%%%%%%%%%%
%%%%%%%%%%%%%%%%%%%%%%%%%%%%%%%%%%%%%%%%%%%%%%%%%
%%%%%%%%%%%%%%%%%%%%%%%%%%%%%%%%%%%%%%%%%%%%%%%%%

\section{Supplementary Material}

\section{Derivation of the kinetic Eq.(4)}

\subsection{Primary asymptotics}

The starting point is the NLS Eq.(1) written in the mode basis, i.e., Eq.(2).
We consider the regime where linear propagation dominates over disorder, which in turn dominates over the nonlinearity. 
Accordingly, we introduce a small dimensionless parameter $\eps$ and we consider the regime $\beta_j \to \beta_j, {\bf C} \to \eps {\bf C}, \gamma \to \eps^2 \gamma$.
For propagation distances of order $\eps^{-2}$, the rescaled mode amplitudes 
$a_j^\eps(z) = a_j(z/\eps^2)$ satisfy
\begin{equation*}
\partial_z a_j^\eps = -i \frac{\beta_j}{\eps^2} a_j^\eps +i \gamma \sum_{l,m,n=0}^{M-1} Q_{jlmn} a_l^\eps a_m^\eps \overline{a_n^\eps} -
\frac{i}{\eps} \sum_{l=0}^{M-1}
C_{jl}(\frac{z}{\eps^2}) a_l^\eps  ,
\end{equation*}
where the bar stands for complex conjugation.
We set 
$c_j^\eps(z) =  a_j^\eps(z)\exp\big( i \frac{\beta_j}{\eps^2} z\big)$. 
The amplitudes $c_j^\eps(z) $ satisfy:
\begin{align}
\partial_z c_j^\eps =
& i \gamma \sum_{l,m,n=0}^{M-1} Q_{jlmn} c_l^\eps c_m^\eps \overline{c_n^\eps} \exp\big( i \frac{\beta_j -\beta_l-\beta_m+\beta_n }{\eps^2} z\big)
\nonumber \\
& -
\frac{i}{\eps} \sum_{l=0}^{M-1}
C_{jl}(\frac{z}{\eps^2}) c_l^\eps \exp\big( i \frac{\beta_j - \beta_l}{\eps^2} z\big)  .
\end{align}
This is the usual diffusion approximation framework \cite{fouque07}.
We get the following result.

\begin{proposition}
\label{prop:1}
The  random process 
$
 ({c}_j^\eps(z) )_{j=0}^{M-1}  
$
converges in distribution in ${\cal C}^0([0,\infty), \mathbb{C}^{M} )$, 
the space of continuous functions from $[0,\infty)$ to $\mathbb{C}^{M}$,
to the Markov process  
$
  (\mathfrak{c}_{j}(z) )_{j=0}^{M-1}   
$
with infinitesimal generator
${\cal L}$:
\begin{align} 
 \label{eq:defL1}
{\cal L}= & {\cal L}_1 + {\cal L}_2+{\cal L}_3 +{\cal L}_4 +{\cal L}_5
,
\end{align}
with
\begin{align*} 
\nonumber {\cal L}_1 = & \frac{1}{2} 
\sum_{j,l=0, j\neq l}^{M-1} 
\Gamma^{\rm OD}_{jl} \big( 
\mathfrak{c}_{j} \overline{\mathfrak{c}_{j}} \partial_{\mathfrak{c}_{l}} \partial_{\overline{\mathfrak{c}_{l}}}
+
\mathfrak{c}_{l} \overline{\mathfrak{c}_{l}} \partial_{\mathfrak{c}_{j}} \partial_{\overline{\mathfrak{c}_{j}}}
\\
&\quad \quad -
\mathfrak{c}_{j}  \mathfrak{c}_{l} \partial_{\mathfrak{c}_{j}} \partial_{ \mathfrak{c}_{l}}
-
\overline{\mathfrak{c}_{j}}  \overline{\mathfrak{c}_{l}} \partial_{\overline{\mathfrak{c}_{j}}} \partial_{ \overline{\mathfrak{c}_{l}}}\big)
 ,
%\label{eq:defL11}
\\
\nonumber 
{\cal L}_2 =& 
 \frac{1}{2} 
\sum_{j,l=0}^{M-1} 
\Gamma^{\rm D}_{j l} \big( 
\mathfrak{c}_{j} \overline{\mathfrak{c}_{l}} \partial_{\mathfrak{c}_{j}} \partial_{\overline{\mathfrak{c}_{l}}}
+
\overline{\mathfrak{c}_{j}} \mathfrak{c}_{l} \partial_{\overline{\mathfrak{c}_{j}}} \partial_{\mathfrak{c}_{l}}
\\
&\quad \quad -
\mathfrak{c}_{j}  \mathfrak{c}_{l} \partial_{\mathfrak{c}_{j}} \partial_{ \mathfrak{c}_{l}}
-
\overline{\mathfrak{c}_{j}}  \overline{\mathfrak{c}_{l}} \partial_{\overline{\mathfrak{c}_{j}}} \partial_{ \overline{\mathfrak{c}_{l}}}\big)  ,
\end{align*}
\begin{align*} 
{\cal L}_3 =& \frac{1}{2} 
\sum_{j=0}^{M-1}  \Gamma^{\rm OD}_{jj} 
\big( \mathfrak{c}_{j} \partial_{\mathfrak{c}_{j}} + \overline{\mathfrak{c}_{j}} \partial_{\overline{\mathfrak{c}_{j}}}
\big)
+i  {\hat \Gamma}^{\rm OD}_{jj}  
\big( \mathfrak{c}_{j} \partial_{\mathfrak{c}_{j}} - \overline{\mathfrak{c}_{j}} \partial_{\overline{\mathfrak{c}_{j}}}
\big) ,  
\\ 
{\cal L}_4 =&  - \frac{1}{2} 
\sum_{j=0}^{M-1}    \Gamma^{\rm D}_{jj} 
\big( \mathfrak{c}_{j} \partial_{\mathfrak{c}_{j}} + \overline{\mathfrak{c}_{j}} \partial_{\overline{\mathfrak{c}_{j}}}
\big) ,\\
{\cal L}_5 =&  i \gamma 
\sum_{l,m,n=0}^{M-1}  \delta^K_{jlmn}  {Q_{jlmn}}
\big( \mathfrak{c}_l  \mathfrak{c}_m  \overline{\mathfrak{c}_n} 
 \partial_{\mathfrak{c}_{j}} - 
 \overline{\mathfrak{c}_l}  \overline{\mathfrak{c}_m}   {\mathfrak{c}_n} \partial_{\overline{\mathfrak{c}_{j}}}
\big)  ,
%\label{eq:defL15}
\end{align*}

where 
$
%\begin{equation}
\delta^K_{jlmn} =  {\bf 1}_{\beta_j-\beta_l-\beta_m+\beta_n=0}
$.
%\end{equation}
In this definition we use the  classical complex derivative:
if $ \zeta=\zeta_r+i\zeta_i$, then $\partial_\zeta=(1/2)(\partial_{\zeta_r}-i \partial_{\zeta_i})$ and
$\partial_{\overline{\zeta}} =(1/2)(\partial_{\zeta_r} +i \partial_{\zeta_i})$,
and the coefficients of the operator ${\cal L}_k$ ($k=1,...,5$) are defined for 
$j ,l= 0, \ldots, M-1$, as follows: 

 \noindent - For all $j \neq l$, $\Gamma_{jl}$ and ${\hat \Gamma}^{\rm OD}_{j l}$ are given by 
\begin{align}
\Gamma^{\rm OD}_{jl} &= 2
\int_0^\infty{\cal R}_{jl}(z) \cos \big((\beta_l-\beta_j)z\big) dz , 
%\label{def:Gammalj}
\\
{\hat \Gamma}^{\rm OD}_{jl} &=
2
\int_0^\infty 
{\cal R}_{jl}(z)  \sin \big( (\beta_l-\beta_j)z \big) dz ,
\end{align}
with ${\cal R}_{jl}(z) $ defined by
\begin{align}
\nonumber
&
{\cal R}_{jl}(z) = \EE [C_{jl}(0)C_{jl}(z)] \\
&= \iint u_j(\br) u_j(\br') \EE[\delta V(0,\br) \delta V(z,\br')]  {u_l}(\br)   {u_l}(\br')d\br d\br'.
\label{def:calRjl}
%{\cal R}_{jl}(z) &:= \EE [C_{jl}(0)C_{jl}(z)] = \iint u_j(\bx) u_j(\bx') \EE[V(0,\bx) V(z,\bx')] \overline{u_l}(\bx)  \overline{u_l}(\bx')d\bx d\bx'.
\end{align}
\noindent - For all $j,l=0,\ldots,M-1$:
\begin{align}
\Gamma^{\rm D}_{jl} =&
\int_0^\infty \EE\big[ C_{jj} (0) C_{ll}(z) \big]   dz
+
\int_0^\infty \EE\big[ C_{ll} (0) C_{jj}(z) \big]   dz .
\end{align}
\noindent 
- For all $j=0,\ldots,M-1$:
\begin{align}
\Gamma^{\rm OD}_{jj} =& -\hspace{-0.05in}\sum_{l =0,l\neq j}^{M-1} \Gamma^{\rm OD}_{jl} ,\quad \quad
{\hat \Gamma}^{\rm OD}_{jj} = -\hspace{-0.05in}\sum_{l =0,l\neq j}^{M-1} {\hat \Gamma}^{\rm OD}_{jl}  .
\end{align} 
\end{proposition}

\subsection{Secondary asymptotics}

We observe that $\Gamma^{\rm OD}$ and ${\hat \Gamma}^{\rm OD}$ depend on the power spectral density of the 
random index perturbation evaluated at the difference of distinct frequencies $\beta_j-\beta_l$, while $\Gamma^{\rm D}$ depends on the power spectral density of the index perturbation evaluated at zero-frequency. 
Therefore, 
when 
%the longitudinal correlation radius $\ell_c$ of the random index perturbation   is larger than 
$L_{lin}=1/\beta_0 \ll \ell_c$, then $\Gamma^{\rm D}$ is larger than $\Gamma^{\rm OD}, {\hat \Gamma}^{\rm OD}$. 
We consider this regime by introducing a small dimensionless parameter $\eta$ with $\Gamma^{\rm D}\to \Gamma^{\rm D}$, $\Gamma^{\rm OD} \to \eta^2 \Gamma^{\rm OD}$, ${\hat \Gamma}^{\rm OD} \to \eta^2 {\hat \Gamma}^{\rm OD}$, $\gamma \to \eta \gamma$.

For propagation distances of order $\eta^{-2}$, we introduce the rescaled mode amplitudes $\mathfrak{c}_j^\eta(z) = \mathfrak{c}_j(z/\eta^2)$.
By Proposition \ref{prop:1}
it is a Markov process with infinitesimal generator
${\cal L}^\eta$:
\begin{equation}
{\cal L}^\eta = 
{\cal L}_1 + \eta^{-2} {\cal L}_2+{\cal L}_3 +\eta^{-2} {\cal L}_4 + \eta^{-1} {\cal L}_5 ,
\label{eq:defL1eta}
\end{equation}
where the operators ${\cal L}_k$ ($k=1,..,5$) are given above.
By (\ref{eq:defL1eta}) the second-order moments satisfy for $j\neq j'$:
\begin{align*}
\partial_z \EE[ \mathfrak{c}_j^\eta \overline{\mathfrak{c}_{j'}^\eta}] = &
-\frac{1}{2\eta^2} (\Gamma^{\rm D}_{jj}+\Gamma^{\rm D}_{j'j'}-2\Gamma^{\rm D}_{jj'}) \EE[ \mathfrak{c}_j^\eta \overline{\mathfrak{c}_{j'}^\eta}]   \\
& \hspace*{-0.3in}
 + \frac{1}{2} \big( \Gamma^{\rm OD}_{jj}+ \Gamma^{\rm OD}_{j'j'}\big) \EE[ \mathfrak{c}_j^\eta \overline{\mathfrak{c}_{j'}^\eta}] 
 +\frac{i}{2}\big( {\hat \Gamma}^{\rm OD}_{jj} - {\hat \Gamma}^{\rm OD}_{j'j'}\big) \EE[ \mathfrak{c}_j^\eta \overline{\mathfrak{c}_{j'}^\eta}] 
 \\
& \hspace*{-0.3in}+ i \frac{\gamma}{\eta} \sum_{l,m,n=0}^{M-1} \delta^K_{jlmn} Q_{jlmn} \EE[ \overline{\mathfrak{c}_{j'}^\eta} \mathfrak{c}_l^\eta  \mathfrak{c}_m^\eta  \overline{\mathfrak{c}_n^\eta}  ] \\
& \hspace*{-0.3in}
- i\frac{\gamma}{\eta}  \sum_{l,m,n=0}^{M-1} \delta^K_{j'lmn} 
{Q_{j'lmn}} 
\EE[{\mathfrak{c}_{j}^\eta} \overline{\mathfrak{c}_l^\eta}  \overline{\mathfrak{c}_m^\eta}   
{\mathfrak{c}_n^\eta}  ]    ,
\end{align*}
up to negligible terms in $\eta$.
Note that 
$
\Gamma^{\rm D}_{jj}+\Gamma^{\rm D}_{j'j'}-2\Gamma^{\rm D}_{jj'} = 
\int_{-\infty}^\infty \EE\big[ (C_{jj} (0) - C_{j'j'}(0) )( C_{jj}(z)-C_{j'j'}(z) )\big]   dz
$
is positive (it is the power spectral density evaluated at $0$ frequency of the stationary process $C_{jj}(z)-C_{j'j'}(z)$ by Bochner's theorem).
%If it is positive (which is the standard case), then 
Therefore $\EE[ \mathfrak{c}_j^\eta \overline{\mathfrak{c}_{j'}^\eta}] $ is exponentially damped and
\begin{equation}
\label{eq:damped2}
\EE[ \mathfrak{c}_j^\eta \overline{\mathfrak{c}_{j'}^\eta}] = O(\eta).
\end{equation}

If $j=j'$, then the mean square amplitudes
$w_j^\eta(z) = \EE[ |\mathfrak{c}_j^\eta(z)|^2] $
satisfy
\begin{align}
\nonumber
\partial_z w_j^\eta  =&
\sum_{l=0,l\neq j}^{M-1} \Gamma_{jl}^{\rm OD} \big(w_l^\eta-w_j^\eta \big)  
\\
&-2 \frac{\gamma}{\eta}   \sum_{l,m,n=0}^{M-1} \delta^K_{jlmn}  Q_{jlmn}  {\rm Im}\Big\{
 \EE[ \overline{\mathfrak{c}_{j}^\eta} \mathfrak{c}_l^\eta  \mathfrak{c}_m^\eta  \overline{\mathfrak{c}_n^\eta}   ] \Big\}  .
 \label{eq:odewj}
\end{align}

By (\ref{eq:defL1eta})  
the fourth-order moments satisfy
\begin{align}
 \partial_z \EE[ \overline{\mathfrak{c}_{j}^\eta} \mathfrak{c}_l^\eta  \mathfrak{c}_m^\eta  \overline{\mathfrak{c}_n^\eta}  ]  =
-\frac{1}{2\eta^2} 
G^{\rm D}_{jlmn}
\EE[ \overline{\mathfrak{c}_{j}^\eta} \mathfrak{c}_l^\eta  \mathfrak{c}_m^\eta  \overline{\mathfrak{c}_n^\eta}  ] 
+ i \frac{\gamma}{\eta} Y_{jlmn}^\eta  \nonumber \\
+  \sum_{j',l',m',n'} M_{jlmn,j'l'm'n'}  \EE[ \overline{\mathfrak{c}_{j'}^\eta}  \mathfrak{c}_{l'}^\eta  {\mathfrak{c}_{m'}^\eta}  \overline{\mathfrak{c}_{n'}^\eta}],
\label{eq:odec4}
\end{align}
up to negligible terms in $\eta$.
The coefficients $G^{\rm D}_{jlmn}$ and the sixth-order moment $Y_{jlmn}^\eta$ are given by
\begin{align}
\nonumber
G^{\rm D}_{jlmn}
=& \Gamma^{\rm D}_{ll}+\Gamma^{\rm D}_{mm}+\Gamma^{\rm D}_{nn}+\Gamma^{\rm D}_{jj}+2\Gamma^{\rm D}_{lm}-2\Gamma^{\rm D}_{ln}\\
& -2\Gamma^{\rm D}_{lj} 
-2\Gamma^{\rm D}_{mn}-2\Gamma^{\rm D}_{mj}+2\Gamma^{\rm D}_{nj},
\label{def:Gammajlmn} \\
\nonumber
Y_{jlmn}^\eta =&    \sum_{l',m',n'=0}^{M-1} \delta^K_{l l'm'n'} S_{ll'm'n'} \EE[ 
  \mathfrak{c}_{l'}^\eta  \mathfrak{c}_{m'}^\eta  \overline{\mathfrak{c}_{n'}^\eta}
  \mathfrak{c}_m^\eta  \overline{\mathfrak{c}_n^\eta}  \overline{\mathfrak{c}_{j}^\eta}]
  \\
&
  \quad  +
  \delta^K_{ml'm'n'} S_{ml'm'n'} \EE[ 
  \mathfrak{c}_l^\eta  \mathfrak{c}_{l'}^\eta  \mathfrak{c}_{m'}^\eta  \overline{\mathfrak{c}_{n'}^\eta}
 \overline{\mathfrak{c}_n^\eta}  \overline{\mathfrak{c}_{j}^\eta}] 
  \nonumber 
  \\
&
 \quad   -  \delta^K_{n l'm'n'}  
{S_{n l'm'n'}}
 \EE[  \mathfrak{c}_l^\eta  \mathfrak{c}_m^\eta  
  \overline{\mathfrak{c}_{l'}^\eta}  \overline{\mathfrak{c}_{m'}^\eta}   
{\mathfrak{c}_{n'}^\eta} \overline{\mathfrak{c}_{j}^\eta} ]
\nonumber \\
& \quad -
\delta^K_{j l'm'n'}  
{Q_{j l'm'n'}} 
\EE[  \mathfrak{c}_l^\eta  \mathfrak{c}_m^\eta  
  \overline{\mathfrak{c}_{n}^\eta}   \overline{\mathfrak{c}_{l'}^\eta}  \overline{\mathfrak{c}_{m'}^\eta}   
{\mathfrak{c}_{n'}^\eta}   ]  ,
\label{def:Yjlmn}
\end{align}
up to negligible terms in $\eta$.
The tensor $M_{jlmn,j'l'm'n'}$ involves the coefficients $\Gamma^{\rm OD}$ and ${\hat \Gamma}^{\rm OD}$.
Note that we have
$ G^{\rm D}_{jlmn}=
\int_{-\infty}^\infty \EE\big[ (C_{ll} (0) + C_{mm}(0) -C_{nn}(0)-C_{jj}(0) )( C_{ll}(z)+C_{mm}(z)-C_{nn}(z)-C_{jj}(z)  )\big]   dz \ge 0
$. 
Therefore, we find from (\ref{eq:odec4}) that 
$$
\EE[ \overline{\mathfrak{c}_{j}^\eta} \mathfrak{c}_l^\eta  \mathfrak{c}_m^\eta  \overline{\mathfrak{c}_n^\eta}  ]  = \frac{2i \eta \gamma}{ G^{\rm D}_{jlmn}} Y_{jlmn}^\eta 
+O(\eta^2)  .
$$
By substituting into (\ref{eq:odewj}) and by using Isserlis formula for the sixth-order moments that appear in the expression (\ref{def:Yjlmn}) of $Y^\eta_{jlmn}$  
we obtain the kinetic Eq.(4):
\begin{align}
&
\partial_z w_j^\eta  = 
\sum_{l=0, l\neq j}^{M-1} \Gamma_{jl}^{\rm OD} \big(w_l^\eta-w_j^\eta \big) 
\nonumber \\
& +
8 \gamma^2   \sum_{l,m,n=0}^{M-1}  \frac{\delta^K_{jlmn}  Q_{jlmn}^2}{G^{\rm D}_{jlmn}} \big(w_l^\eta w_m^\eta w_j^\eta + w_l^\eta w_m^\eta w_n^\eta 
\nonumber \\
& \hspace*{1.5in} -  w_j^\eta w_n^\eta w_m^\eta - w_j^\eta w_n^\eta w_l^\eta  \big)  .
\label{eq:kineq2}
\end{align}
The second term in (\ref{eq:kineq2}) has a form analogous to the conventional collision term of the wave turbulence kinetic equation \cite{zakharov92}. Exploiting the invariances properties of the tensors $Q_{jlmn}$ and $G^{\rm D}_{jlmn}$, as well as the property $G^{\rm D}_{jlmn} \ge 0$, it can be shown that the collision term conserves the particle number $N=\sum_j w_j$, the energy $E=\sum_j \beta_j w_j$, and exhibits a $H-$theorem of entropy growth $\partial_z S(z) \ge 0$, where the nonequilibrium entropy reads $S(z)=\sum_j \log[w_j(z)]$ (note that, for simplicity we omitted to write  the superscript $\eta$). The entropy growth saturates at thermal equilibrium. The RJ equilibrium distribution that maximizes the entropy $S[w_j]$, under the constraints that $N$ and $E$ are conserved, reads 
\begin{align}
w_j^{\rm RJ}=T/(\beta_j - \mu),
\label{eq:rj}
\end{align} 
where $1/T$ and $-\mu/T$ are the Lagrange multipliers associated to the conservation of $E$ and $N$.
There is a one to one relation relation between the pair $(N,E)$ and $(T,\mu)$: 
The values of the conserved quantities $(N,E)$ determine uniquely $(T,\mu)$, and thus the RJ equilibrium (\ref{eq:rj}).
%The collision term then describes an irreversible evolution to the RJ equilibrium distribution.}

\subsection{Degenerate modes}
In this section we assume that the modes may be degenerate.
% and we revisit the previous results.
The detailed derivation of the kinetic equation accounting for mode degeneracy is cumbersome and will be reported elsewhere.
Here we report the main results.

There are $G$ distinct wavenumbers:
$$
\{ \beta^{(g)} ,\, g=1,\ldots, G\} ,
$$
and the mode indices can be partitioned into $G$ groups ${\cal G}^{(g)}$, $g=1,\ldots,G$:
$$
{\cal G}^{(g)} = \{ p=1,\ldots,N ,\, \beta_p = \beta^{(g)}\} .
$$

We obtain the kinetic equation
\begin{align*}
\nonumber
\partial_z w^{(g)}= 
8 \gamma^2
 \sum_{g_1,g_2,g_3=1}^G  \delta^{(gg_1g_2g_3)}
 q^{(gg_1g_2g_3)} \big( w^{(g)} w^{(g_3)} w^{(g_2)} \\
 +w^{(g)} w^{(g_3)} w^{(g_1)}
- w^{(g_1)} w^{(g_2)} w^{(g)}  - w^{(g_1)} w^{(g_2)} w^{(g_3)} \big)   ,
 \label{eq:wg2}
\end{align*}
where
\begin{equation*}
q^{(gg_1g_2g_3)}=\frac{1}{|{\cal G}^{(g)}|} 
\sum_{j \in {\cal G}^{(g)} , l\in {\cal G}^{(g_1)}, m \in {\cal G}^{(g_2)}, n\in {\cal G}^{(g_3)} }
\hspace*{-0.3in}
  Q_{jlmn} {Q}^{(gg_1g_2g_3)}_{jlmn}  
%\Big]    .
\end{equation*}
where
\begin{align*}
\nonumber
& {\bm Q}^{(gg_1g_2g_3)} =\big( {Q}^{(gg_1g_2g_3)}_{jlmn} \big)_{j\in {\cal G}^{(g)},
l\in {\cal G}^{(g_1)}, m \in {\cal G}^{(g_2)}, n\in {\cal G}^{(g_3)} } \\
&=
({\bf M}^{(gg_1g_2g_3)})^{-1} \big(  ( 
{Q_{jlmn}} 
)_{ j\in {\cal G}^{(g)},
l\in {\cal G}^{(g_1)}, m \in {\cal G}^{(g_2)}, n\in {\cal G}^{(g_3)} }  \big)  .
\end{align*}
The tensor ${\bf M}^{(gg_1g_2g_3)}$ 
(seen as a $q \times q$ matrix with $q=|{\cal G}^{(g)}| |{\cal G}^{(g_1)}| |{\cal G}^{(g_2)}| |{\cal G}^{(g_3)}| $) is given by
\begin{align*}
& \sum_{ j' \in {\cal G}^{(g)}, l'\in {\cal G}^{(g_1)}, m' \in {\cal G}^{(g_2)}, n'\in {\cal G}^{(g_3)}   }
M^{(gg_1g_2g_3)}_{jlmn,j'l'm'n'}
 w_{j'l'm'n'}   \\
&  =
   \sum_{l'\in {\cal G}^{(g_1)},m'\in {\cal G}^{(g_2)}} 2\gamma_{ll'mm'} 
   w_{jl'm'n}  \\
&\quad 
 +  \sum_{n'\in {\cal G}^{(g_3)},j'\in {\cal G}^{(g)} } 2\gamma_{nn'jj'} 
   w_{j'lmn'}   \\
&\quad  - \sum_{l'\in {\cal G}^{(g_1)},n'\in {\cal G}^{(g_3)}} 2\gamma_{ll'nn'} 
  w_{jl'mn'}   \\
&\quad 
 - \sum_{l'\in {\cal G}^{(g_1)},j'\in {\cal G}^{(g)}} 2\gamma_{ll'jj'} 
   w_{j'l'mn}  \\
& \quad- \sum_{m'\in {\cal G}^{(g_2)},n'\in {\cal G}^{(g_3)}} 2\gamma_{mm'nn'} 
    w_{jlm'n'}    \\
&\quad 
     - \sum_{m'\in {\cal G}^{(g_2)},j' \in {\cal G}^{(g)}} 2\gamma_{mm'jj'} 
    w_{j'lm'n}  \\
&\quad  +  \sum_{l',l''\in {\cal G}^{(g_1)}} \gamma_{l''l'll''} 
    w_{jl'mn}    
+  \sum_{m',m''\in {\cal G}^{(g_2)}} \gamma_{m''m'mm''} 
  w_{jlm'n} \\
&\quad +  \sum_{n',n''\in {\cal G}^{(g_3)}} \gamma_{n''n'nn''} 
   w_{jlm n'}
+  \sum_{j',j''\in {\cal G}^{(g)}} \gamma_{j''j'jj''} 
   w_{j' l m n}    .
\end{align*}
where 
$$
\gamma_{pqp'q'}
=2
\int_0^\infty \EE\big[ C_{pq}(z)C_{p'q'}(0)\big] e^{i (\beta_p-\beta_q)z} dz .
$$

\subsection{Numerical simulations}

%Solving the NLS Eq.(??) (main text) becomes challenging when the number of modes increases because of the large number of terms involved in the sums of the nonlinear term.
%We implemented a split-step method where the nonlinear term is computed in real space $\br=(x,y)$, while the linear diffraction and disorder effects are solved in the modal basis $\{u_p(\br)\}$. 
%The integrals involved in the projections of the field $\psi(\br)$ into the modal basis are optimized by a Gaussian quadrature \cite{laegsgaard17}.
%%with a grid of 128$^2$ points to minimize the time cost. 

\noindent
{\it Implementation of disorder:}
To implement the disorder in the simulations of the NLS Eq.(2), we considered an {\it exact discretization} of the Ornstein-Uhlenbeck process.
The propagation axis is divided in intervals with deterministic lengths $\Delta z$, with $\Delta z < \ell_c$.
% smaller than $l_\beta$ (say, $\Delta z = l_\beta/10$).
The random function $\mu(z)$ is stepwise constant over each elementary interval
%${\bf D}_p(z) =  \sum_{j=1}^3 \nu_{p,j,k} \bsigma_j$    if 
$z \in [k\Delta z, (k+1)\Delta z)$,   
%\mbox{ for } p=1,\ldots,N_*
%$
%if $z \in [k\Delta z, (k+1)\Delta z) $
where $\mu_{0}\sim {\cal N}(0,\sigma^2/2)$ denotes the Gaussian distribution, 
%for  $p=1,\ldots,N_*$, $j=1,2,3$, and 
$
\mu_{k} = \sqrt{1-2\Delta z/\ell_c}\mu_{k-1} +  
\sqrt{2\Delta z/\ell_c} {\cal N}( 0,\sigma^2/2) , 
$
with ${\cal N}( 0,\sigma^2/2)$ all independent and identically distributed.
%We recall that 
%${\cal N}( 0,\frac{\sigma_\beta^2}{2})$ denotes the Gaussian distribution $f(x)=\exp(- x^2 / \sigma_\beta^2) / \sqrt{\pi \sigma_\beta^2}$.

\medskip

\noindent
{\it Model of disorder:}
We have considered in the numerical simulations an elliptical parabolic potential $V(\bx)=q_x x^2 + q_y y^2$, with $u_{p_x,p_y}(x,y) = \sqrt{\kappa_x \kappa_y} (\pi p_x! \, p_y! \, 2^{p_x+p_y})^{-1/2} \, H_{p_x}(\kappa_x x) \, H_{p_y}(\kappa_y y) \, \exp[-(\kappa_x^2 x^2+\kappa_y^2 y^2)/2]$ the normalized Hermite-Gaussian functions with corresponding eigenvalues $\beta_p=\beta_{p_x,p_y}=\beta_{0x}(p_x+1/2)+\beta_{0y}(p_y+1/2)$, with $\kappa_x= (q_x/\alpha)^{1/4}$, $\kappa_y= (q_y/\alpha)^{1/4}$, $\beta_{0x} = 2\sqrt{\alpha q_x}$, $\beta_{0y} = 2\sqrt{\alpha q_y}$, and the radii of the fundamental mode $r_{0x}=1/\kappa_x=\sqrt{2\alpha/\beta_{0x}}$, $r_{0y}=1/\kappa_y=\sqrt{2\alpha/\beta_{0y}}$.
%In the following we may consider $\beta_{0y}=\beta_{0x}/2$. Then, if we denote by $V_{0,m}$ the depth of the potential, and $V_{0,m}$ is slightly larger than $\beta_{0x}(p_{x,m}+1/2)$ for some $p_{x,m}$, then $p_x=0,1... p_{x,m}$ and $p_y=0,1... p_{y,m}$ with $p_{y,m}=2p_{x,m}$. 
%However, as will be discussed below, it is preferable to chose $\beta_{0x}/\beta_{0y}$ irrational.
%The total number of modes is given by the number of distinct pairs $(p_x,p_y)$ such that $\beta_{p_x,p_y} \le V_{0,m}$.\\
%Remark: When applied to the 1st model of disorder ($\nu=2$), the elliptical parabolic fiber still gives $\Gamma_{jlmn}=0$ even in 2D.\\

We have considered the following form of model of disorder:
$\delta V(\bx,z)=\mu(z) \cos(\kappa_x b_x x) \cos(\kappa_y b_y y)$, with $\EE[\mu(0) \mu(z)]= \sigma^2 f(z)$, $f(z)=\exp(-|z|/\ell_c)$. The advantage of this model is that the matrices ${\bf C}, {\bf \Gamma}^{\rm D}, {\bf \Gamma}^{\rm OD}$ can be computed in analytical form.
We have $C_{nk}(z)=\mu(z) C_{n_x k_x}^0 C_{n_y k_y}^0=\mu(z) \int u_{n_x}(x) \cos(\kappa_x b_x x) u_{k_x}(x) dx$ $\times  \int u_{n_y}(y) \cos(\kappa_y b_y y) u_{k_y}(y) dy$.\\
%Then $C_{nk}(z)$ factorizes into the $x$ and $y$ components.\\
% which are computed independently of each other by directly applying the 1D results.\\
Then we have for $j_{x},j_y,l_x, l_y \geq 0$: $C_{j,j+2l}=\mu(z) C_{j_x,j_x+2l_x}^0 C_{j_y,j_y+2l_y}^0$ where we denote for $s=x$ or $s=y$:\\
\begin{align*}
C_{j_s,j_s+2l_s}^0= (-1)^{l_s} b_s^{2l_s} \exp(-b_s^2/4) \quad \quad \quad  \\ 
\quad \quad \quad \times L_{j_s}^{2l_s}(b_s^2/2) \frac{\sqrt{j_s!/(j_s+2l_s)!}}{2^{l_s}}
\end{align*}
and $C_{j_s,j_s+2l_s+1}^0=0$, where $L_{j}^{l}$ is the generalized Laguerre poynomial \cite[formula 7.388.7]{grad}.
%Note that $C_{j_x,j_x+2l_x}^0$ is computed by only considering positive values of $l_x \geq 0$, and the corresponding negative values are obtained by using the fact that $C_{n_x,k_x}^0=C_{k_x,n_x}^0$.\\
In particular 
$C_{j_s j_s}^0= \exp(-b_s^2/4) L_{j_s}(b_s^2/2)$, where $L_{j}$ is the  Laguerre poynomial. 
For $j_{x},j_y,l_x, l_y \geq 0$, 
we have $\Gamma^{\rm D}_{jl}=2\sigma^2 \ell_c  C_{j_x j_x}^0 C_{j_y j_y}^0 C_{l_x l_x}^0 C_{l_y l_y}^0$.\\
For $n_x,k_x,n_y,k_y \geq 0$ we obtain:
$$
\Gamma^{\rm OD}_{n,k}=\frac{2\sigma^2 \ell_c {\cal R}_{n_x,k_x}^0 {\cal R}_{n_y,k_y}^0}{1+\ell_c^2 [\beta_{0x}(n_x-k_x)+\beta_{0y}(n_y-k_y)]^2},
$$ 
%The matrices ${\cal R}_{n_x,k_x}^0$ are given  
%for $j_{x},j_y,l_x, l_y \geq 0$: 
%${\cal R}_{j,j+2l}(z)= \sigma^2 f(z) {\cal R}_{j_x,j_x+2l_x}^0 {\cal R}_{j_y,j_y+2l_y}^0$ 
where ${\cal R}_{j_s,j_s+2l_s+1}^0=0$ and 
$$
{\cal R}_{j_s,j_s+2l_s}^0=b_s^{4l_s} \exp(-b_s^2/2) L_{j_s}^{2l_s}(b_s^2/2)^2 ( j_s!/(j_s+2l_s)!) 2^{-2l_s}.
$$ 

\medskip 

\noindent
%{\it Parameters used in the simulations:} 
In order to avoid high values of $\Gamma^{\rm OD}_{n,k}$, we have considered an irrational ratio $\beta_{0x}/\beta_{0y}=\sqrt{2}$, so that $\beta_{0x}(n_x-k_x)+\beta_{0y}(n_y-k_y) \neq 0$.
Parameters are ($b_x = 0.4, b_y = 0.5$) in Figs.~1-2, and ($b_x = 0.4, b_y = 0.3$) in Fig.~3. In all cases we considered $M=$46 modes.
The value of $L_{nl}=1/(\gamma N/A_{eff}^0)$ in the simulations is computed by considering that all the power $N$ is in the fundamental mode of effective area $A_{eff}^0=1/\int |u_0|^4(\br) d\br$.
%${\cal L}^{\rm RJ}_{\rm kin}$ given in Eq.(8) (main text) involves in the denominator the tensor $Q_{jlmn}$, whose zeros lead to a divergence of ${\cal L}^{\rm RJ}_{\rm kin}$. 
%A threshold value for $Q_{jlmn}$ has been adjusted in such a way that ${\cal L}^{\rm RJ}_{\rm kin}$ properly describes the thermalization length scale observed in the simulations of the kinetic Eq.(\ref{eq:kineq2}).

\begin{figure}
\includegraphics[width=.9\columnwidth]{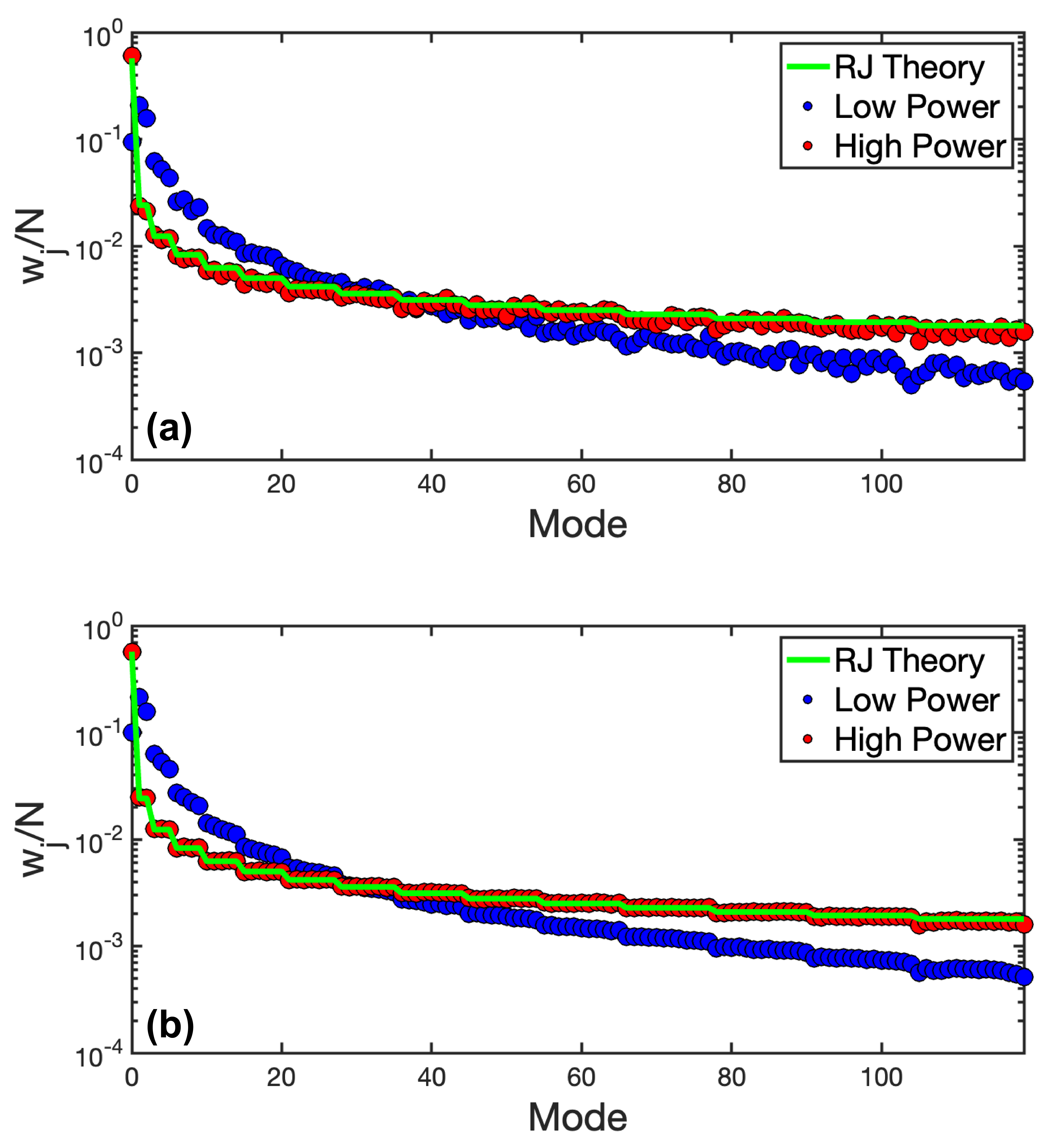}
\caption{
\baselineskip 8pt
{\bf Observation of RJ thermalization (without disorder):} 
Experimental modal distributions $w_j/N$ (circles), for an individual realization of the launched speckle beam (a), for an average over the realizations of speckle beams (b).  
The red circles report the results at high power ($N=7$kW, nonlinear regime), and the blue circles at low power ($N=0.23$kW, linear regime).
The condensate fraction is $w_0/N = 0.6$ for $E/N = 1.94 \times 10^4$m$^{-1}$ (a); 
$w_0/N = 0.57$ for $E/N = 2.05 \times 10^4$m$^{-1}$ (b).
Corresponding theoretical RJ equilibrium distribution $w_j^{\rm RJ}/N$ given from Eq.(\ref{eq:rj}) (green line): The quantitative agreement with the experimental data (red circles) is obtained without using any adjustable parameter.
}
\label{fig:supp_rj} 
\end{figure}

\section{Experimental methods}

\noindent
{\bf 1) Setup:} 
The experimental setup has been described in detail in Ref.\cite{PRL20}.
Here we summarize the main characteristics.
The source is a Nd:YAG laser delivering subnanosecond pulses (400ps) at $\lambda_0=$1064 nm. 
%To prevent unwanted feedback into the laser cavity, we use an optical isolator and 
We control the power with a half-wave plate and a polarizer. 
The laser beam was collimated and passed through a glass diffuser plate placed in the vicinity of the Fourier plane of a 4f-optical system.
The beam was launched into the MMF. 
%by using a lens $f_1$ = 15mm. 
The near-field (NF) and far-field (FF) intensity distributions are measured at the fiber output following the procedure of Ref.\cite{PRL20}.  
%The near-field (NF) intensity distribution of the output beam was magnified and imaged on a first CCD camera owing to a two lens telescope optical system.
% with $f_2$ = 8 mm and $f_3$ = 150 mm. 
%The CCD camera was placed on a rail orthogonal to the beam propagation in order to remove or put the camera back on the beam path. 
%The far-field (FF) intensity distribution of such a magnified image was obtained by placing it in the object focal-plan of a lens 
%$f_4$ = 250 mm 
%and using a second CCD camera positioned in its image (Fourier) focal-plan.
We used a 12m-long graded-index MMF whose refractive index profile exhibits a parabolic shape in the fiber core with a maximum core index (at the center) of $n_{\rm co}$=1.470 and $n_{\rm cl}=1.457$ for the cladding at the pump wavelength of 1064nm (numerical aperture NA=0.195, fiber radius $R=26 \mu$m, $\beta_0 \simeq 5\times 10^3$m$^{-1}$).
%The corresponding parabolic potential reads $V(\br)=q |\br|^2$ for $|\br| \le R$, where $R=26 \mu$m is the fiber radius and $q=k_0(n_{\rm co}^2-n_{\rm cl}^2)/(2 n_{\rm co} R^2)$, $k_0=2\pi/\lambda_0$ the laser wave-number.
The MMF guides $M \simeq 120$ modes.
% (i.e. $g=15$ groups of non-degenerate modes).
The truncation of the potential introduces a frequency cut-off in the FF spectrum 
$k_c=(2\pi/\lambda_0)\sqrt{n_{\rm co}^2-n_{\rm cl}^2}$.
For details, see Supplementary Methods in Ref.\cite{PRL20}. 

The temporal spectrum was controlled by an optical spectrum
analyzer (OSA) (600 to 1700nm range). 
%We verified that the NF and FF intensity distributions at the fiber output are not significantly affected by the presence of the Raman effect. 
The spectral analysis showed that the power scattered by self-stimulated Raman effect is in average $\sim$5\% of the injected power. 
Also, the spectral analysis did not
reveal the presence of parametric lines that would be induced
by coupling between dispersive and nonlinear effects.
\\

%Recalling that $\beta_0=2\sqrt{\alpha q}$, we have 
%$k_c=\sqrt{2V_0/\beta_0}/r_0$.
%with $g=V_0/\beta_0$ is the number of groups of non-degenerate modes ($M=g(g+1)/2$).
%Eq.(\ref{eq:k_c}) is also obtained from the numerical aperture of the fiber  NA=$\sin \beta=\sqrt{n_{\rm co}^2-n_{\rm cl}^2}=k_c/k_0$.
%The MMF has a core, a cladding (with radius 62.5$\mu$m), and a highly absorbing polymer-coating with refractive index larger than the core.
%Then leaky radiation modes in the cladding are rapidly absorbed during propagation due to their large penetration in the polymer-coating: we measured a typical absorption length of $\sim 15$cm.
%Since this length is much smaller than that required to generate leaky modes, the nonlinear excitation of leaky modes can be neglected.

%The temporal spectrum was controlled by an optical spectrum analyzer (OSA) (600 to 1700nm range). We verified that the NF anf FF intensity distributions at the fiber output are not significantly affected by the presence of the Raman effect. The spectral analysis showed that the power scattered by self-stimulated Raman effect is in average $\sim 5\%$ of the injected power. 
%Moreover the spectral analysis did not reveal the presence of parametric lines that would be induced by coupling between dispersive and nonlinear effects.
%\\

\begin{figure}
\includegraphics[width=.9\columnwidth]{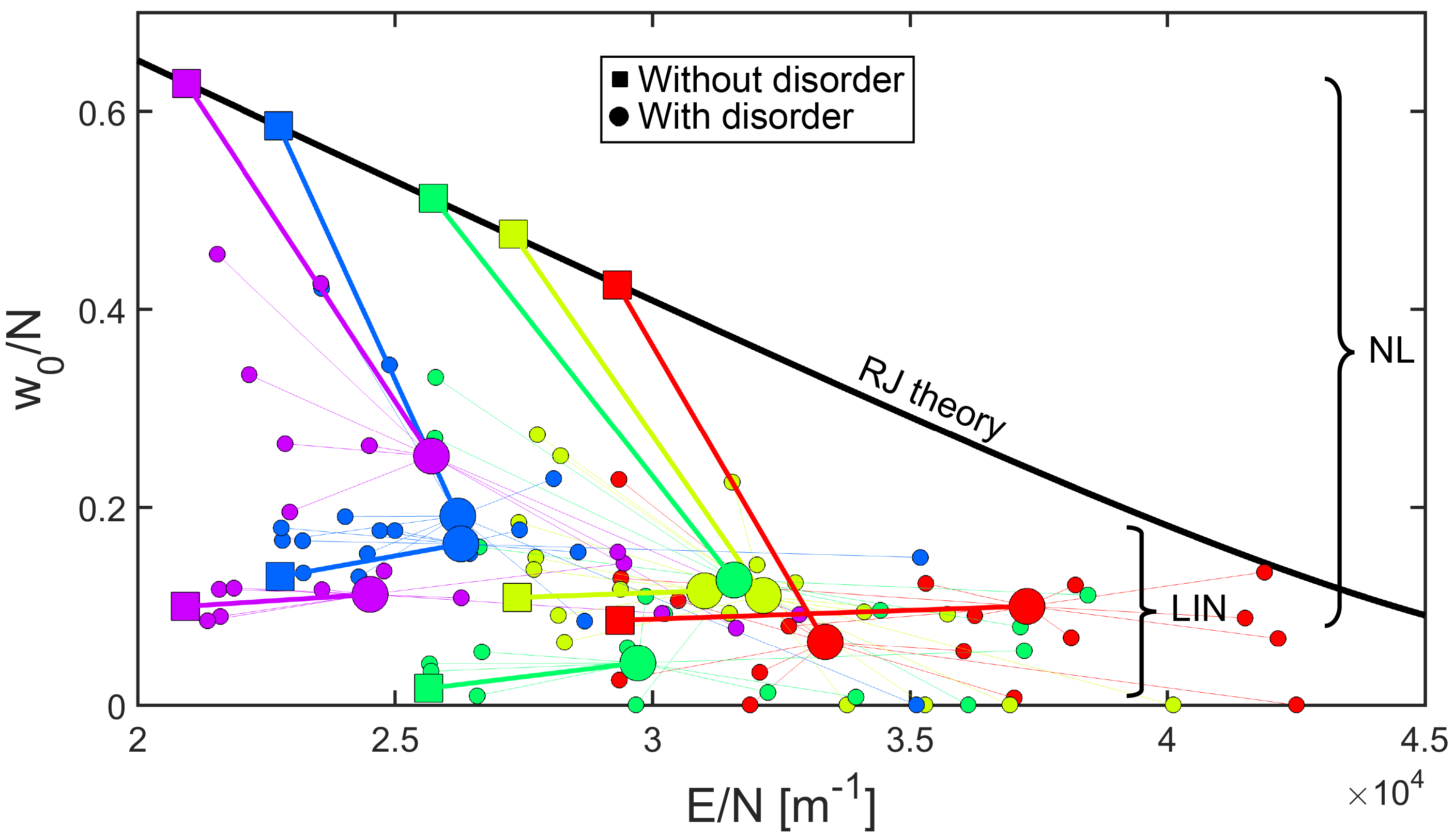}
\caption{
\baselineskip 8pt
{\bf Suppression of light thermalization and condensation by strong disorder:} 
Measurements of the condensate fraction $w_0/N$ vs energy $E/N$ at small power (linear (LIN) regime) and high power (nonlinear (NL) regime), for a large strength of random mode coupling corresponding to an increase of the energy due to disorder of $\Delta \overline{E/N} \simeq 19$\%.
%40\% of power losses.  
The black solid line reports the condensate fraction from the RJ theory, $w_0^{\rm RJ}/N$ vs $E/N$.
In the absence of strong disorder (squares): $w_0/N$ increases as the power increases, and reaches the value predicted by the RJ theory (solid line) -- each color refers to a different value of the energy $E/N$.
In the presence of strong disorder (big circles): the energy $E/N$ increases (the squares are shifted to the big circles of the same color).
%The small circles report 10 different realizations of disorder for each color, the big circles report the corresponding average over realizations.
The big circles report the average over 10 different realizations of disorder (10 small circles for each color).
At variance with Fig.~4, here the strength of random mode coupling is so large that RJ thermalization and condensation are inhibited by strong diorder.
}
\label{fig:supp_exp} 
\end{figure}

\noindent
{\bf 2) Conservation of power $N$ and energy $E$ during propagation without strong disorder:} 
The conservation of the power has been verified by keeping fixed the conditions of injection of the speckle beam into the MMF: We measured $N$ at the fiber output, and then at the input by cutting the fiber at $20$cm, and we always obtained a relative power difference less than 1\%.
The conservation of the energy requires the NF and FF intensity measurements, which provide the potential energy $E_{\rm pot}=  \int V(\br) |\psi(\br)|^2 d\br$, and the kinetic energy $E_{\rm kin}= \int \alpha |\nabla\psi|^2 d\br$, with $E=E_{\rm kin}+E_{\rm pot}$.
%We have already discussed the conservation of the power $N$. 
%measured that the power is conserved to ??\% during the propagation through the fiber, which is consistent with the power loss coefficient of ??m$^{-1}$.
%The verification of the conservation of the energy density $E/N$ is more delicate due to the population of higher order modes during the propagation of the beam through the MMF.
%As for the power, we compare the ``input" and ``output" measurements for fixed fiber launch conditions:
%are known to affect the number of modes (i.e. energy $E$) excited at the fiber input. 
%Specifically, the conservation of $E/N$ has been verified in two steps.
The energy $E_{\rm out}$ is measured at the fiber output at $L=12$m.
Without altering the fiber launch conditions, the fiber is cut to $20$cm to get $E_{\rm in}$.
%The measurements of $E_{\rm in}$ and $E_{\rm out}$ then refer to an individual realization of the speckle beam (without average over the realizations).
The procedure is repeated for different speckle beams (i.e., for different values of the energy $E$), by moving the diffuser before injection into the MMF.
We always obtained $|E_{\rm out}-E_{\rm in}|/E_{\rm moy} < 1$\% for values of the energy that span the range of the condensation curve, i.e. $w_0^{\rm RJ}/N$ varying from 0 to 0.7.
%Note that we have also checked experimentally the equipartition of the kinetic contribution and the potential contribution to the total energy $\left<E_{\rm kin}\right> = \left<E_{\rm pot}\right> = E/2$ (by averaging over the realizations), see Supplementary Methods in Ref.\cite{PRL20}. 
\\

\noindent
{\bf 3) Experimental observation of RJ thermalization:}
In the absence of strong disorder (i.e., absence of applied stress induced on the fiber), we observe the process of thermalization to the RJ equilibrium distribution, $w_j^{\rm RJ}=T/(\beta_j-\mu)$.
In the experiments, the modal populations ($w_j$) are computed by using the Gerchberg-Saxton algorithm, which allows us to retrieve the transverse phase profile of the field from the NF and the FF intensity distributions measured in the experiments \cite{fienup82}.
By projecting the complex field over the modes of the MMF (Gauss-Hermite basis) we get the complete modal distribution $w_j/N$, $j=0,1,..,M-1$. 
A typical example  is reported in Fig.~\ref{fig:supp_rj} showing the modal distribution $w_j/N$ recorded experimentally at low-power (linear regime) and high-power (nonlinear regime), and its comparison to the RJ equilibrium distribution.
Fig.~\ref{fig:supp_rj}(a) reports a single realization of the speckle beam, Fig.~\ref{fig:supp_rj}(b) reports an average over 60 realizations of speckle beams.
The quantitative agreement between the experimental results and the theoretical RJ distribution is obtained without using adjustable parameters.
\\

%\newpage

\begin{figure}
\includegraphics[width=.9\columnwidth]{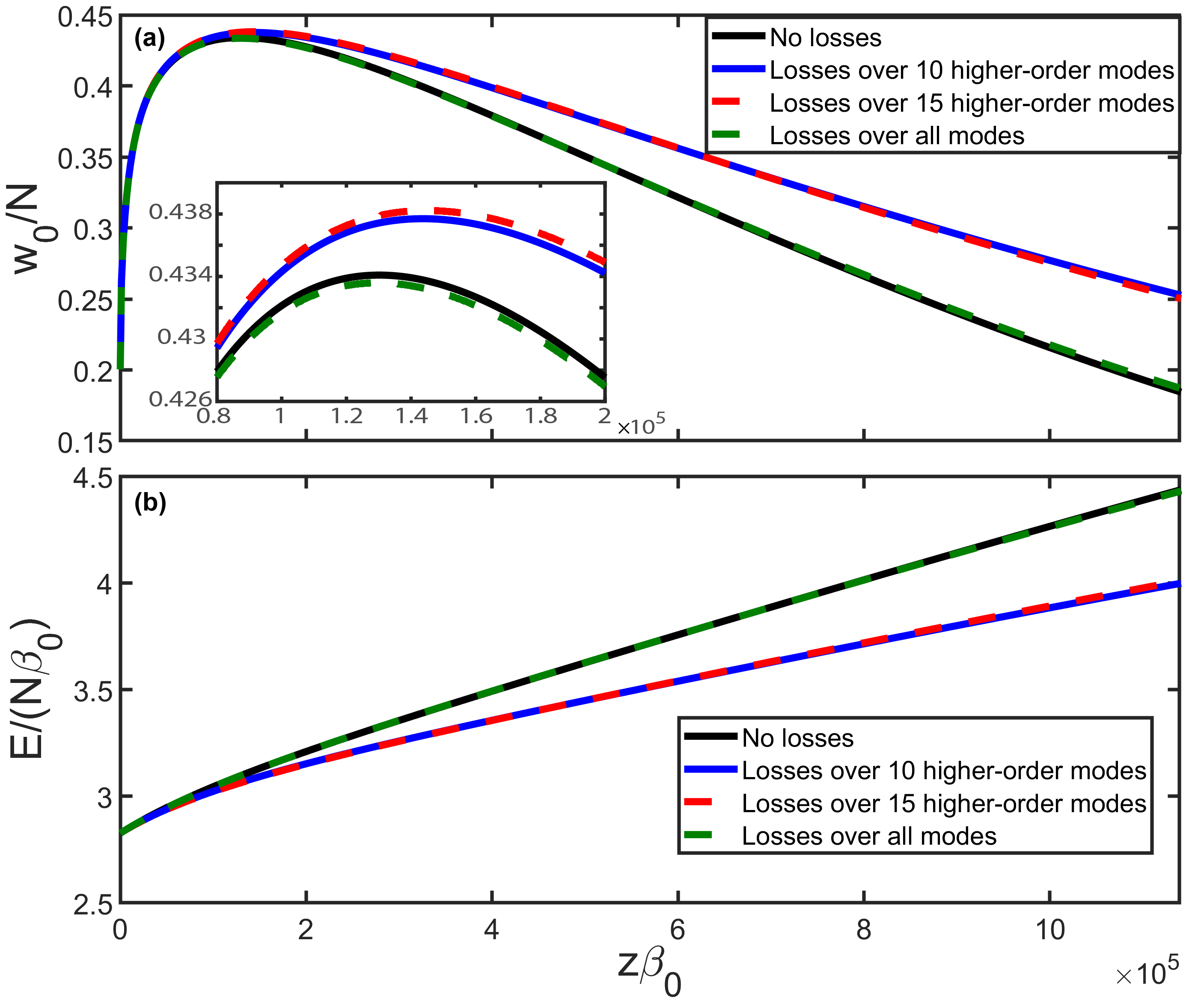}
\caption{
\baselineskip 8pt
{\bf Impact of losses on the condensate fraction:} 
Simulation of the kinetic Eq.(\ref{eq:kineq2}) for the same parameters as Fig.~2: 
In the absence of losses (black), and when 10\% of losses are distributed among all modes (dashed green), among the higher-order 10 modes of the fiber (blue), among the higher-order 15 modes of the fiber (dashed red).
Condensate fraction $w_0(z)/N(z)$ vs $z$, where $N(z)$ is the local value of the power accounting for the losses (a), and corresponding evolutions of the energy $E(z)/(N(z) \beta_0)$ (b).
The inset in (a) shows a zoom: The condensate peak relevant to the experiments is only weakly affected by the presence of the losses.
}
\label{fig:simul} 
\end{figure}

\noindent
{\bf 4) Experimental procedure with strong disorder (applied stress):} 
The laser beam is passed through a diffuser before injection of the speckle beam into the MMF. 
The coupling conditions and the position of the diffuser then fix the energy density $E/N$ of the speckle beam.
In the absence of applied stress, $E/N$ is conserved through propagation in the MMF (see point 2) above).
We report in Fig.~4(a), 5 different ensembles of measurements, each one corresponding to a fixed position of the diffuser (i.e. fixed value of the energy $E/N$ without applied stress). 
For a given fixed position of the diffuser, we perform the following steps i)-vii) to retrieve 10 different realizations of disorder in Fig.~4(a):\\
{\bf i)} Without applying any stress, we measure the NF and FF intensity patterns at high power ($N=7$kW, nonlinear regime), and compute $E/N$ and $w_0/N$ (squares in Fig.~4(a)). We verify that $w_0/N$ is in agreement with the value predicted by the RJ theory, see Ref.\cite{PRL20} for details.\\
{\bf ii)}  At low power ($N=0.23$kW, linear regime) we measure the NF and FF intensity patterns and compute $E/N$ and $w_0/N$.\\
{\bf iii)} We return to the previous higher power ($N=7$kW, nonlinear regime) and we verify that we recover the same NF speckle beam as in step i).\\
{\bf iv)} Then we apply stress to a specific location of the MMF.
The stress is applied by using clamps mounted on a linear translation manual stage whose position is controlled at the micrometer scale.
We adjust the amount of stress by measuring the power losses (10\% in Fig.~4(a), corresponding to $\Delta \overline{E/N} \simeq 6$\%). 
Once the stress is adjusted, the power is increased up to the same average power of step i).
We then measure the NF and FF intensity patterns and compute $E/N$ and $w_0/N$ (small circles in Fig.~4(a)).\\
{\bf v)} In a next step we decrease the power ($N=0.23$kW, linear regime), we measure the NF and FF intensity patterns and compute $E/N$ and $w_0/N$ (small circles in Fig.~4(a)).\\
{\bf vi)} We return to the previous higher power ($N=7$kW, nonlinear regime) and remove the applied stress. 
We verify that we recover the same initial NF speckle beam as in step i).\\
{\bf vii)} We repeat the steps iv)-v)-vi) 10 times to get 10 different realizations of strong disorder (small circles). Each disorder realization is achieved by applying stress to a different position of the MMF by rotating the drum on which it is wound.

\smallskip
\noindent
The procedure i)-vii) is repeated for a larger amount of applied stress (disorder), corresponding to an increase of energy due to disorder of $\Delta \overline{E/N} \simeq 11$\% in Fig.~4(b) (20\% of power losses), and $\Delta \overline{E/N} \simeq 19$\% in Fig.~\ref{fig:supp_exp} (40\% of power losses).
In Fig.~\ref{fig:supp_exp} the strength of random mode coupling is so large that RJ thermalization and condensation are inhibited by strong disorder.

\smallskip
\noindent
Note that losses induced by strong disorder only weakly affect the condensate fraction through the propagation in the MMF, as illustrated in the simulation reported in Fig.~\ref{fig:simul}.
We have considered 10\% of losses (over the propagation length $z \beta_0=11 \times 10^5$), in the case where losses are distributed homogeneously in mode space, and non-homogeneously in mode space (only the higher-order modes experience losses).
We have considered the parameters of the simulation reported in Fig.~2, which refers to the most interesting regime where linear disorder effects and nonlinear effects are of the same order, ${\cal L}^{\rm RJ}_{\rm kin} \lesssim {\cal L}^{\rm eq}_{\rm kin}$.
The condensate peak relevant to the experiments is only weakly affected by the losses, see the inset in Fig.~\ref{fig:simul}(a).
Note that, for larger propagation lengths, the losses concentrated on the higher-order modes  reduce the effective number of modes and thus limit the increase of energy $E/(N \beta_0)$ due to disorder (Fig.~\ref{fig:simul}(b)), which in turn leads to an increase of the condensate fraction (Fig.~\ref{fig:simul}(a)).

%\end{widetext}
%\newpage


\begin{thebibliography}{99}
\bibitem{comment}
More exactly, ${\cal L}^{\rm eq}_{\rm kin}= 1/\lambda_2^{\rm OD}$ where $-\lambda_2^{\rm OD}$ is the second eigenvalue of the matrix $\tilde{\boldsymbol{\Gamma}}^{\rm OD}$, with $\tilde{\Gamma}_{jl}^{\rm OD}=\Gamma^{\rm OD}_{jl}$ for $j \neq l$ and $\tilde{\Gamma}^{\rm OD}_{jj}=-\sum_{l \neq j}\Gamma^{\rm OD}_{jl}$, the first eigenvalue being $\lambda_1^{\rm OD}=0$.
\end{thebibliography}

\end{document}